\documentclass[twoside,twocolumn,9pt]{article}

\usepackage{extsizes}
\usepackage[super,sort&compress,comma]{natbib} 
\usepackage[version=4]{mhchem}
\usepackage[left=1.5cm, right=1.5cm, top=1.785cm, bottom=2.0cm]{geometry}
\usepackage{balance}
\usepackage{mathptmx}
\usepackage{sectsty}
\usepackage{graphicx} 
\usepackage{lastpage}
\usepackage[format=plain,justification=justified,singlelinecheck=false,font={stretch=1.125,small,sf},labelfont=bf,labelsep=space]{caption}
\usepackage{float}
\usepackage{fancyhdr}
\usepackage{fnpos}
\usepackage[english]{babel}
\addto{\captionsenglish}{%
  
}
\usepackage{array}
\usepackage{droidsans}
\usepackage{charter}
\usepackage[utf8]{inputenc}
\usepackage[T1]{fontenc}
\usepackage[usenames,dvipsnames]{xcolor}
\usepackage{setspace}
\usepackage[compact]{titlesec}
\usepackage{hyperref}

\usepackage{breakurl} 

\usepackage{xurl}

\usepackage{siunitx}
\usepackage{textcomp}
\usepackage{stfloats}
\usepackage{amssymb} 
\usepackage{orcidlink}

\newsavebox{\foobox}
\newcommand{\slantbox}[2][0]{\mbox{%
    \sbox{\foobox}{#2}%
    \hskip\wd\foobox
    \pdfsave
    \pdfsetmatrix{1 0 #1 1}%
    \llap{\usebox{\foobox}}%
    \pdfrestore
}}
\newcommand\unslant[2][-.2]{\slantbox[#1]{#2}}

\usepackage{epstopdf}

\definecolor{cream}{RGB}{222,217,201}

\begin{document}

\pagestyle{fancy}
\thispagestyle{plain}
\fancypagestyle{plain}{
\renewcommand{\headrulewidth}{0pt}
}

\makeFNbottom
\makeatletter
\renewcommand\LARGE{\@setfontsize\LARGE{15pt}{17}}
\renewcommand\Large{\@setfontsize\Large{12pt}{14}}
\renewcommand\large{\@setfontsize\large{10pt}{12}}
\renewcommand\footnotesize{\@setfontsize\footnotesize{7pt}{10}}
\makeatother

\renewcommand{\thefootnote}{\fnsymbol{footnote}}
\renewcommand\footnoterule{\vspace*{1pt}%
\color{cream}\hrule width 3.5in height 0.4pt \color{black}\vspace*{5pt}} 
\setcounter{secnumdepth}{5}

\makeatletter 
\renewcommand\@biblabel[1]{#1}            
\renewcommand\@makefntext[1]%
{\noindent\makebox[0pt][r]{\@thefnmark\,}#1}
\makeatother 
\renewcommand{\figurename}{\small{Fig.}~}
\sectionfont{\sffamily\Large}
\subsectionfont{\normalsize}
\subsubsectionfont{\bf}
\setstretch{1.125} 
\setlength{\skip\footins}{0.8cm}
\setlength{\footnotesep}{0.25cm}
\setlength{\jot}{10pt}
\titlespacing*{\section}{0pt}{4pt}{4pt}
\titlespacing*{\subsection}{0pt}{15pt}{1pt}

\fancyfoot{}
\fancyfoot[LO,RE]{\vspace{-7.1pt}}
\fancyfoot[CO]{\vspace{-7.1pt}\hspace{13.2cm}}
\fancyfoot[CE]{\vspace{-7.2pt}\hspace{-14.2cm}}
\fancyfoot[RO]{\footnotesize{\sffamily{ ~\textbar  \hspace{2pt}\thepage}}}
\fancyfoot[LE]{\footnotesize{\sffamily{\thepage~\textbar\hspace{3.45cm} }}}
\fancyhead{}
\renewcommand{\headrulewidth}{0pt} 
\renewcommand{\footrulewidth}{0pt}
\setlength{\arrayrulewidth}{1pt}
\setlength{\columnsep}{6.5mm}
\setlength\bibsep{1pt}

\makeatletter 
\newlength{\figrulesep} 
\setlength{\figrulesep}{0.5\textfloatsep} 

\newcommand{\topfigrule}{\vspace*{-1pt}%
\noindent{\color{cream}\rule[-\figrulesep]{\columnwidth}{1.5pt}} }

\newcommand{\botfigrule}{\vspace*{-2pt}%
\noindent{\color{cream}\rule[\figrulesep]{\columnwidth}{1.5pt}} }

\newcommand{\dblfigrule}{\vspace*{-1pt}%
\noindent{\color{cream}\rule[-\figrulesep]{\textwidth}{1.5pt}} }

\makeatother

\twocolumn[
  \begin{@twocolumnfalse}
\sffamily

\noindent\LARGE\centering{\textbf{\textit{In situ} and \textit{operando} laboratory X-ray absorption spectroscopy at high temperature and controlled gas atmosphere with a plug-flow fixed-bed cell}} 
\vspace{0.3cm}

\begin{tabular}{m{0.5cm} p{16.5cm} }

 & \noindent\large{Sebastian Praetz,\orcidlink{0000-0001-6852-7616}$^{\ast}$\textit{$^{a\ddag}$} Emiliano Dal Molin,\orcidlink{0000-0002-1377-3999}\textit{$^{b\ddag}$} Delf Kober,\textit{$^{b}$} Marko Tesic,\orcidlink{0009-0004-2615-1138}$^{a}$ Christopher Schlesiger,\orcidlink{0000-0002-9407-4267}$^{a}$ Peter Kraus,\orcidlink{0000-0002-4359-5003}\textit{$^{c}$} Julian T. Müller,\orcidlink{0000-0003-0703-8573}\textit{$^{b}$} Jyothilakshmi Ravi Aswin,\textit{$^{d}$} Daniel Grötzsch,$^{a}$ Maged F. Bekheet,\orcidlink{0000-0003-1778-0288}\textit{$^{b}$} Albert Gili,\orcidlink{0000-0001-7944-7881}\textit{$^{e}$} Aleksander Gurlo,\orcidlink{0000-0001-7047-666X}\textit{$^{b}$} and Birgit Kanngießer$^{a}$}\\
 & \noindent\large{$\ddag$~These authors contributed equally to this work}\\
 & \noindent\large{${\ast}$ Electronic email: sebastian.praetz@tu-berlin.de}\\

& \noindent\normalsize{The capabilities of a plug-flow fixed-bed cell for \textit{operando} studies of heterogeneous catalysts are demonstrated using laboratory-based X-ray absorption spectroscopy (XAS) with a von Hámos spectrometer. The cell operates at temperatures up to \SI{1000}{\celsius} and pressures up to \SI{10}{bar}, equipped with three mass flow controllers and two infrared lamps for rapid heating under inert/reactive gas atmospheres. Proof-of-principle studies include \textit{in situ} \ce{MnO} oxidation in 5\,\% \ce{Ni/MnO} and \textit{operando} Ni nanoparticle evolution in 20-NiO/COK-12 (\SI{20.2}{\percent} \ce{NiO} on \ce{SiO2}) during \ce{CO2} methanation before/after activation. Within 5--15\,min per spectrum, oxidation state changes are resolved while catalytic activity is simultaneously quantified by online GC. Extended datasets and methods are available in the ancillary file SI.pdf (or Supplementary Information file). A shortened version of this work has been published as a Technical Note in \textit{Journal of Analytical Atomic Spectrometry} (2026, \textbf{41}, 1208--1211, DOI: \href{https://doi.org/10.1039/D6JA00027D}{10.1039/D6JA00027D}). This manuscript provides an extended version including additional datasets, analysis, and methodological details beyond the published article.}

\end{tabular}

 \end{@twocolumnfalse} \vspace{0.6cm}

]
  


\renewcommand*\rmdefault{bch}\normalfont\upshape
\rmfamily
\section*{}
\vspace{-1cm}


\footnotetext{\textit{$^{a}$~Technische Universität Berlin, Institute of Physics and Astronomy, Hardenbergstraße 36, 10623 Berlin, Germany. E-mail: sebastian.praetz@tu-berlin.de}}
\footnotetext{\textit{$^{b}$~Technische Universität Berlin, Faculty III Process Sciences, Institute of Materials Science and Technology, Chair of Advanced Ceramic Materials, Hardenbergstr. 40, 10623 Berlin, Germany.}}
\footnotetext{\textit{$^{c}$~Technische Universität Berlin, Conductivity and Catalysis Lab, Hardenbergstr. 40, 10623 Berlin, Germany.}}
\footnotetext{\textit{$^{d}$~Helmholtz-Zentrum Berlin, Department Atomic-Scale Dynamics in Light-Energy Conversion, 14109 Berlin, Germany.}}
\footnotetext{\textit{$^{e}$~Helmholtz-Zentrum Berlin für Materialien und Energie, 14109 Berlin, Germany.}}






\section{\label{sec:introduction}Introduction}
Heterogeneous catalysts play a crucial role in various applications, including the industrial manufacturing of chemicals,\cite{CIRIMINNA2021,kumar_applications_2024} automotive and industrial emission control\cite{CHUNG2025} (\textit{e.g.}, reduction of harmful emissions such as CO and \ce{NO_x} from vehicles),\cite{DEY2020} and the conversion of biomass into value-added products.\cite{ABDULLAH2017} In the context of renewable energy and decarbonization, they are particularly important for the synthesis of methane and methanol,\cite{Ren2023,Darji2023, Gili2025} enabling the efficient and selective conversion of carbon-containing feedstocks into valuable chemicals and fuels. In methanation, catalysts such as Ni-containing materials facilitate the hydrogenation of CO and \ce{CO2} into methane, a key process in the production of synthetic natural gas (SNG) and carbon recycling. These catalytic processes are essential for sustainable energy applications and the development of green chemical production pathways.\cite{kumar_applications_2024}

A comprehensive understanding of catalytic systems can be achieved by combining electron-based techniques, such as electron microscopy, with photon-based methods, including X-ray diffraction (XRD), X-ray photoelectron spectroscopy (XPS) and X-ray absorption spectroscopy (XAS).\cite{Bergmann2019} By correlating physical and chemical properties with catalytic performance -- evaluated in terms of activity, selectivity, and stability -- it is possible to gain insights into reaction mechanisms, active sites, and deactivation pathways.\cite{Gili2018, Bonmasser2020} Establishing structure-activity relationships is essential for the rational design and optimization of heterogeneous catalysts.

XAS, also referred to as X-ray absorption fine structure spectroscopy (XAFS), is typically categorized into two energy regions: X-ray absorption near-edge structure (XANES), which focuses on the pre-edge, edge, and close post-edge energies, and extended X-ray absorption fine structure (EXAFS), which deals with energies in the far post-edge region. While XANES can provide information on oxidation state, electronic structure, and coordination geometry, EXAFS enables the analysis of local atomic environments, including bond lengths, coordination numbers, and structural disorder. This well-established and highly versatile method is widely used to study the electronic structure of atoms, as well as the coordination and bonding distances between them. XAS found its origin in the laboratory in 1913 and 1920 by de Broglie,\cite{broglie_recherches_1913} Fricke\cite{fricke_k-characteristic_1920} and Hertz\cite{hertz_uber_1920} and is frequently used at synchrotron radiation facilities. In the last decade, this method has also gained traction in laboratory environments with the advent of advanced laboratory spectrometers.\cite{Malzer2021, Zimmermann2020, Seidler2014, Schlesiger2015, Zoltan2016, Honkanen2019}

The laboratory spectrometer used in this work has already demonstrated its capability to measure \textit{ex situ} in the fields of catalysis research,\cite{Le2017, Dimitrakopoulou2018,Oliveira2020,Bekheet2021,Oliveira2023, Ebert2024, Gili2024} materials science\cite{Zhao2019, Menezes2019, Wang2021, Yang2025} and others.\cite{Motz2023, Bauer2024, Praetz-Schlesiger2025} Shown by recent works of Kallio \textit{et al.}\cite{Kallio2022, Genz2022, Johansen2024} using the HelXAS lab spectrometer,\cite{Honkanen2019} and Praetz \textit{et al.}\cite{Praetz-Groetzsch2025,Praetz-Johanson2025} with the spectrometer described in this work, laboratory XAS spectrometers are not limited to \textit{ex situ} measurement. Importantly, while previous \textit{operando} studies with HelXAS were restricted to XANES, the present setup enables \textit{operando} measurements that extend partially into the EXAFS region at higher temperatures. The scan-free approach and broad spectral bandwidth of this spectrometer allow simultaneous acquisition over a wide energy range, significantly reducing measurement times while maintaining comparable signal-to-noise ratios (SNR) compared to previous \textit{in situ} and \textit{operando} studies.

Due to the dynamic nature of catalytic systems, \textit{in situ} and \textit{operando} XAS measurements are crucial for understanding structural and electronic changes under working conditions. Comprehensive reviews highlight recent advances in laboratory-based XAS instrumentation driven by high-brightness microfocus X-ray sources with sub-10\,\unslant{µ}m spot sizes,\cite{Yun2016,Tuohimaa2008,Bartzsch_2017,Malzer2021} enabling spatially resolved XAFS mapping, as well as by liquid-metal-jet (MetalJet) sources providing up to two orders of magnitude higher flux than conventional rotating anodes through continuous microjet targets,\cite{Hemberg2003,Zimmermann2020} allowing XANES acquisition on sub-minute timescales. These source developments are complemented by advances in crystal optics, particularly Highly Annealed Pyrolytic Graphite (HAPG) mosaic crystals, and compact Rowland-circle spectrometer designs.\cite{Schlesiger2015,Schlesiger2020,Praetz-Groetzsch2025} Together, these innovations have enabled laboratory-based \textit{operando} measurements on minute timescales, suitable for monitoring catalyst activation, reduction, and other dynamic processes, thereby reducing reliance on synchrotron facilities while positioning laboratory cell–spectrometer combinations as complementary tools for time-resolved catalysis studies.\cite{Zimmermann2020,Malzer2021}

These recent advances in laboratory-based XAS instrumentation have made it possible to perform time-resolved \textit{operando} measurements with scan durations in the range of 5 to 15 minutes.\cite{Praetz-Groetzsch2025,Praetz-Johanson2025,Kallio2022} Although, this time resolution is not sufficient to capture ultra-fast dynamics accessible by quick XAFS (QXAFS),\cite{Sekizawa2017,Pasquini2021} it is sufficient to monitor key processes such as catalyst activation, reduction, re-oxidation, or deactivation that occur on the timescale of minutes to hours.\cite{Shalabi1997,Loewert2020,Sato2022} In particular, oxidation state changes of transition metals (TM) like nickel (Ni) or manganese (Mn) during reduction in hydrogen or under reaction conditions can be clearly observed, offering valuable insights into the active state of the catalyst.

At elevated temperatures, however, the interpretation of \textit{in situ} and \textit{operando} XAS data requires particular care, as thermal effects such as Debye–Waller damping can significantly reduce EXAFS amplitudes and subtly modify XANES features. These temperature-dependent effects can bias quantitative analyses, especially when high-temperature spectra are compared to reference data acquired at room temperature (RT). Consequently, the use of temperature-matched reference spectra is essential for reliable interpretation of oxidation states and local structure under reaction conditions.

In this work, the oxidation state changes of heterogeneous catalysts are investigated by measuring the absorption K-edge of different 3d-elements (Mn and Ni) to demonstrate a proof of principle, specifically following the oxidation of MnO to \ce{Mn2O3} and tracking the oxidation state of Ni during both catalyst activation and \ce{CO2} methanation. The measurements are performed under \textit{in situ} and \textit{operando} conditions using a slightly modified plug-flow fixed-bed cell reactor by Bischoff \textit{et al.}\cite{Bischoff2024} 
To demonstrate the \textit{operando} lab-XAS performance under catalytic conditions, \ce{NiO} nanoparticles on silica (\ce{SiO2}) (20-NiO/COK-12) (see Section~\ref{sec:20-Ni-COK-12}) at the Ni K-edge were investigated, which are a common catalyst for \ce{CO2} methanation (see Eq.~(\ref{eqn:methanation})).\cite{RONSCH2016} 

\begin{equation}
\label{eqn:methanation}
    \ce{CO2 + 4H2 <=> CH4 + 2H2O } 
\end{equation}

\ce{CO2} methanation is usually in competition with a secondary reaction, known as the "reverse water-gas shift" (RWGS) which produces CO. CO is the thermodynamically preferred product at lower pressures, and higher temperatures.\cite{Zhang2021}

\begin{equation}
\label{eqn:rwgs}
    \ce{CO2 + H2 <=> CO + H2O } 
\end{equation}

The use of nickel catalysts for both of these reactions is widespread, and it is well known that the NiO present must be reduced to Ni to become active. This reaction is typically studied between 300\,°C and 400\,°C, and at pressures ranging from 1 to 10\,bar.\cite{Medina2025} We used this well-studied system to demonstrate the cell's capabilities to measure these changes and measure the appearance of catalytic activity by quantifying the products in the outflow.
Parts of this work have been published in condensed form as a Technical Note in \textit{Journal of Analytical Atomic Spectrometry}.\cite{PraetzDalMolin2026} The present manuscript provides a substantially extended version including additional datasets, analysis, and methodological details.
The main focus of this work is to demonstrate and discuss the challenges, capabilities and limitations of the spectrometer for performing \textit{in situ}/\textit{operando} measurement with this cell.




\begin{figure*}[t]
 \centering
 \includegraphics[width=75mm,trim = 230 90 250 30,clip]{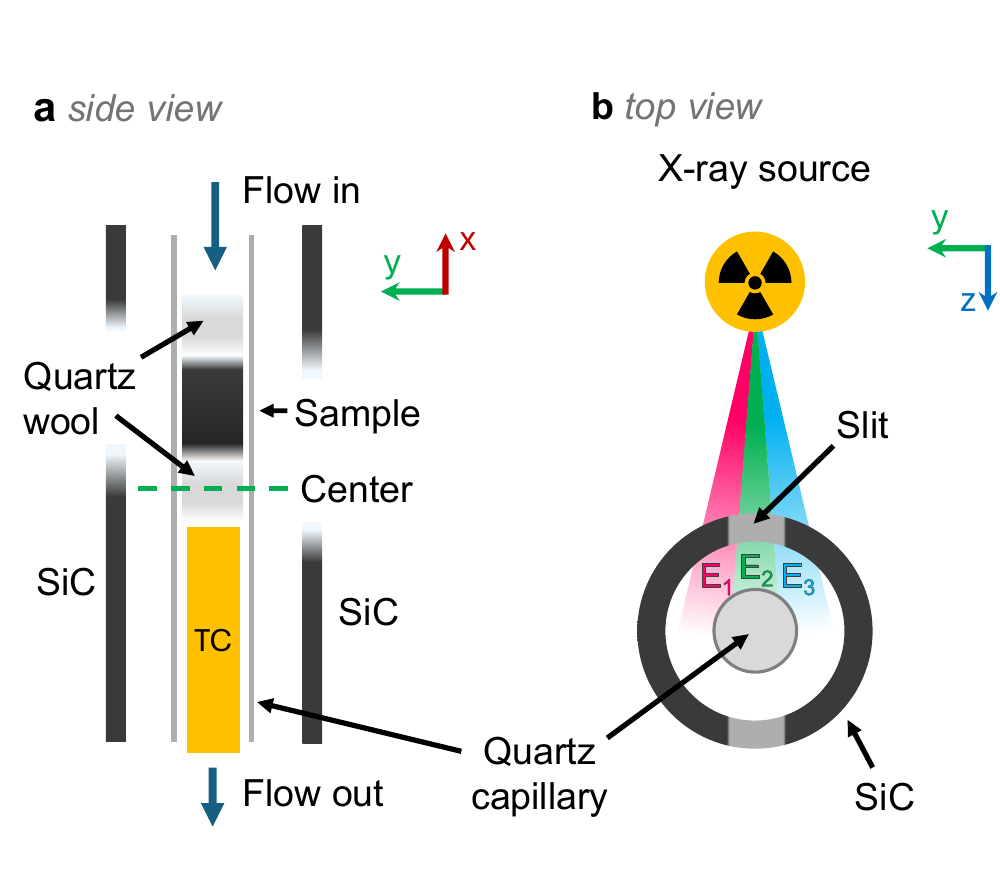}
 \hspace{10mm}
  \includegraphics[width=90mm,trim = 140 50 200 90,clip]{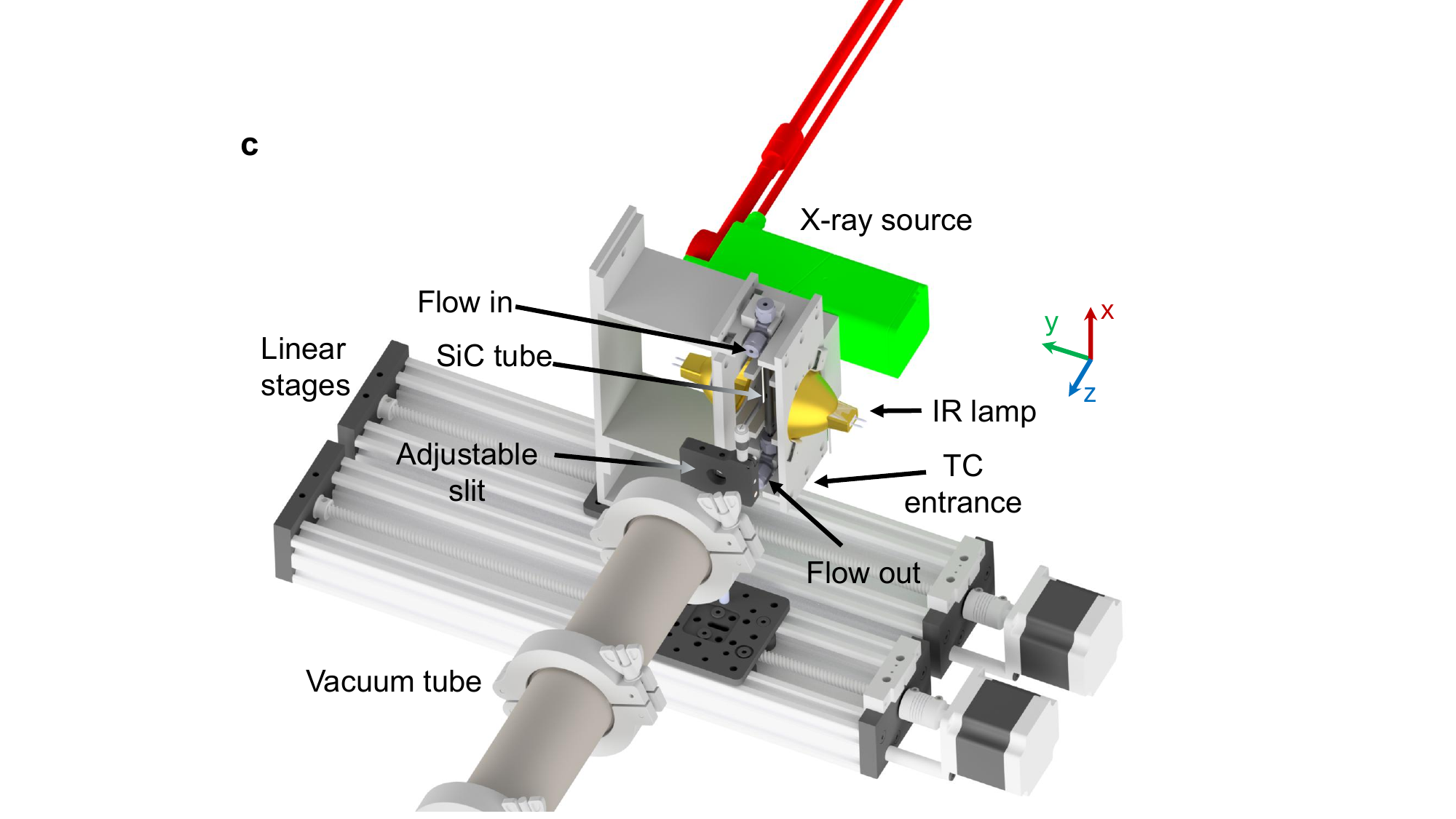}
 \caption{\textbf{a}: Schematic view of sample, quartz wool and TC position inside the capillary surrounded by the SiC tube. \textbf{b}: Top view of the  capillary in the SiC tube showing the X-ray beam path. \textbf{c}: CAD model to the orientation of the IR tube furnace in the von Hámos set up.}
 \label{fig:cell}
\end{figure*}

\section{\label{sec:exp}Experimental Section}
Section~\ref{sec:von-Hamos} outlines the von Hámos setup implemented in this study, with particular emphasis on the challenges associated with the use of an \textit{in situ}/\textit{operando} cell with capillaries, which is explained further in Section~\ref{sec:IR-furnace}. Section~\ref{sec:GC} describes the gas chromatograph (GC) employed to evaluate the catalytic activity. Finally, Section~\ref{sec:samples-references} provides an overview of the investigated catalysts and their synthesis procedures, along with additional samples and references.


\subsection{\label{sec:von-Hamos}XAS setup}
The laboratory XAS spectrometer used in this work is a self-developed wavelength-dispersive spectrometer in von Hámos geometry.\cite{Schlesiger2015,Schlesiger2020} In this setup, a cylindrically curved crystal diffracts and focuses the X-rays emitted from the X-ray tube onto a position-sensitive detector, with both the incident beam and the detector aligned along the cylinder axis.\cite{v_hamos_rontgenspektroskopie_1932} The spectrometer is equipped with a micro-focus X-ray tube MCBI 50B-70 Mo optimised to 15 kV (30\,W at 15\,kV, rtw RÖNTGEN-TECHNIK DR. WARRIKHOFF GmbH \& Co. KG), a curved, highly annealed pyrolytic graphite (HAPG) mosaic crystal -- optimized for XANES measurements with short acquisition times -- and a hybrid photon counting CMOS detector (EIGER2 R 500K) with 512 × 1030 pixels and
a pixel size of 75\,\unslant{µ}m × 75\,\unslant{µ}m. The sample is measured in transmission and placed in close proximity in front of the X-ray source (see Fig.~\ref{fig:cell}(c) and Fig.~S1 in the SI). The X-ray tube is optimized to reach 30\,W at 15\,kV. To lower the absorption in the air and increase the efficiency of the spectrometer, the beam path between the components was evacuated with vacuum tubes (5\,mbar pressure inside) with 25\,\unslant{µ}m Kapton$^\textnormal{\textregistered}$ windows (see Fig.~\ref{fig:cell}(c) and Fig.~S1 in the SI).  

In the von Hámos geometry, different photon energies are spatially separated along the horizontal (meridional) plane and therefore traverse different regions of the sample before being projected onto the detector, while the vertical (sagittal) plane does not contribute to energy discrimination. This geometrical characteristic highlights the importance of homogeneous sample conditions, ideally within a low-conversion (differential) kinetic regime, to minimize gradients in composition or oxidation state across the probed volume. Deviations from a flat and uniform sample geometry -- such as curved or uneven surfaces -- can therefore lead to energy-dependent distortions in the measured XAFS spectra. An extended discussion of these effects can be found in the PhD thesis of Christopher Schlesiger.\cite{schlesiger-rontgenabsorptionsspektroskopie-2019}
The simultaneously measured sample area depends on the investigated energy range and typically covers a few square millimeters.\cite{Schlesiger2015} The measured sample area decreases with increasing focus energy.\cite{schlesiger-rontgenabsorptionsspektroskopie-2019} Samples, when measuring \textit{ex situ}, are prepared by applying the sample as powder on adhesive tape (about 1\,cm × 1\,cm sample area) or as Ø13\,mm pellet, with the option of adding a binder to dilute the sample to achieve an optimal thickness for transmission measurement. A detailed description of the sample preparation can be found in Praetz and Schlesiger \textit{et al.}\cite{Praetz-Schlesiger2025} If the sample size is not covering the whole illuminated (or usable) area in the sample plane, parts of the spectrum will be cut-off as it is the case for thin capillaries, see Fig.~\ref{fig:cell}(b). The capillaries (from CM Scientific, Ireland) mainly used in the experiments in this work have an outer diameter of 1.0\,mm and an inner diameter of 0.8\,mm (see Section~\ref{sec:IR-furnace}), and are positioned vertically within the setup. While this orientation will introduce distortions in the resulting spectrum due to the capillary geometry, placing the capillary horizontally would reduce the sample area contributing in the sagittal plane, thereby decreasing the signal.

To acquire the absorption spectrum as $\mu (E) Q$ (where $\mu$ is the mass attenuation coefficient, $Q$ is the mass deposition of the analyte and $E$ the photon energy in eV), two measurements are necessary: one with the sample ($I_\textnormal{d}$) and one without ($I_0$), the so-called empty measurement. For capillary measurements, $I_0$ is recorded without an empty capillary. Since the capillary used does not cover the entire measurable sample area -- which is further limited by the SiC tube with a 1\,mm slit (see Section \ref{sec:IR-furnace}) -- the use of an adjustable slit is essential (see Fig.~\ref{fig:cell}(c)). Due to the mosaic structure there is no clear spatial cut-off due to a given slit. Therefore, the intensity that is cut-off due to the SiC tube slit but not in the $I_0$ measurement will have a significant overlap with the unrestricted intensity in the image plane leading to strongly distorted spectra. In usual cases the influence of the small sample is barely visible at all in the resulting spectra.
The slit opening is selected according to the capillary diameter and its position to the slit, so that only the material within the capillary, with an extent of 0.8\,mm, is measured. Additionally, the slit is aligned prior to positioning the infrared (IR) furnace setup in order to select the desired energy range and to facilitate the alignment of the furnace with the SiC tube and the capillary inside.

In the following Section~\ref{sec:overview}, the measurable spectral ranges of various absorption edges are presented, along with the impact of the geometry deviation of the capillary from a flat, homogeneous sample.

\begin{figure*}[h]
\centering
    \includegraphics[width=180mm,trim = 0 0 0 0]{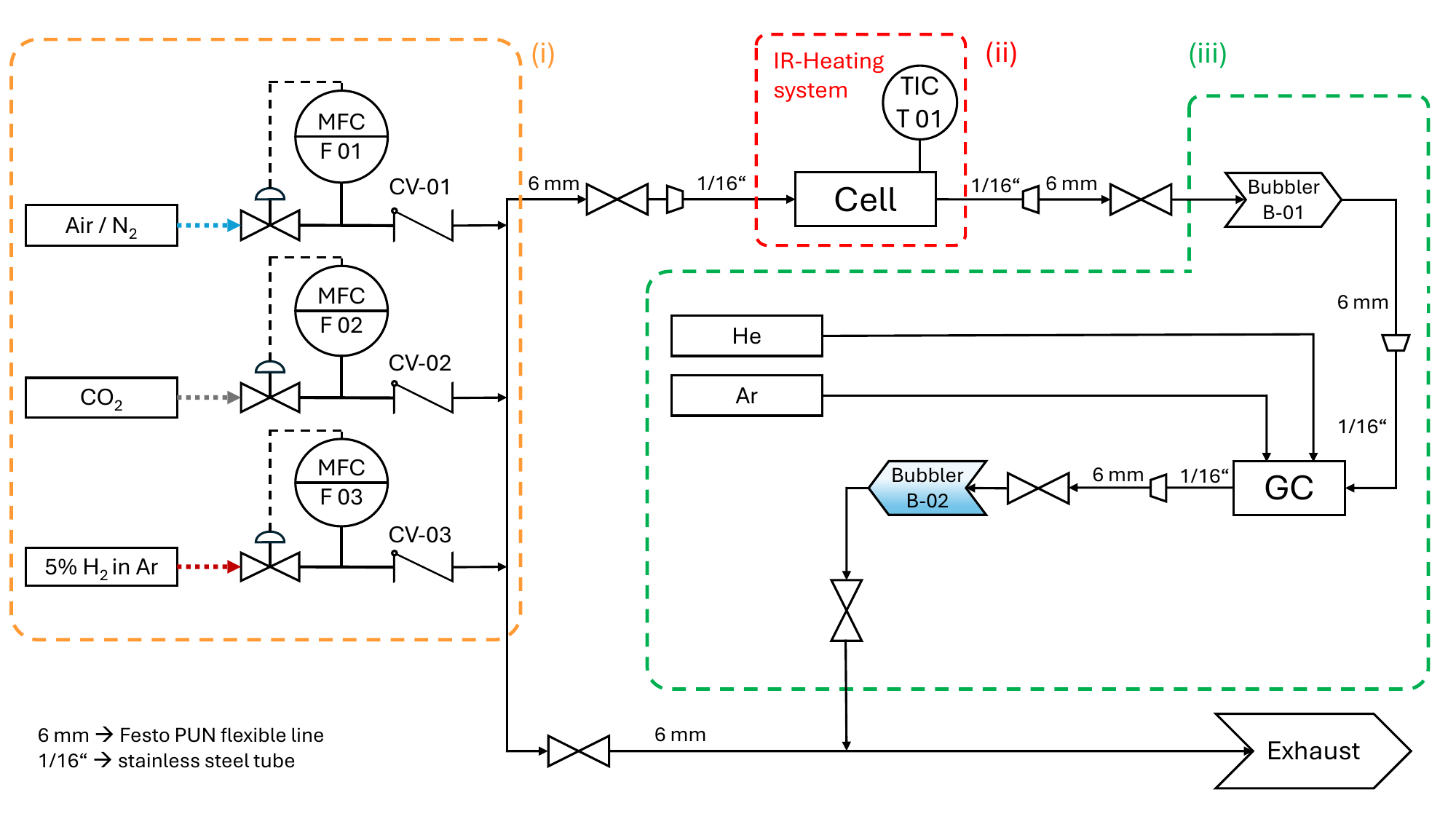}
    \caption{P\&ID of the \textit{In situ}/\textit{operando} reactor setup. With (i) gas flow control unit, (ii) reaction cell and (iii) gas analysis unit. MFC: mass flow controller; CV: check valve; TIC: temperature indicator and controller; GC: Gas chromatograph. Bubbler B-01 is empty to trap water during reaction. Bubbler-B02 contains water to indicate the gas flow. Optional: a pressure control unit (PC) can be installed between the reaction cell and Bubbler B-01.}
    \label{fig:flow-chart}
\end{figure*}


\subsection{\label{sec:IR-furnace}IR-furnace reactor cell setup}
The reactor for the \textit{in situ} and \textit{operando} XAS measurements used in this work was previously demonstrated by Bischoff \textit{et al.}~\cite{Bischoff2024} in synchrotron-based XAFS and XRD studies. For adaptation to the laboratory-scale von Hámos setup (see Section~\ref{sec:von-Hamos}), several modifications were implemented, as outlined below. A schematic overview of the full gas flow and instrumentation layout (P\&ID) is shown in Fig.~\ref{fig:flow-chart}. The setup comprises three main components: (i) the upstream gas flow control system, (ii) the reaction cell and (iii) the downstream gas analysis unit. The reactor cell (ii) can also be used solely with the upstream gas flow control system (i), or entirely on its own if neither gas analysis nor gas flow is required.

\subsubsection*{(i) Upstream: Gas flow control}
The upstream section (i) (see Fig.~\ref{fig:flow-chart}) consists of three EL-Flow Prestige mass flow controllers (MFCs 01–03; Bronkhorst, Netherlands) connected to gas cylinders, for this study containing 5\,\% \ce{H2} in Ar (Arcal 15, \ce{H2} (5\,\%\,±\,0.5\,\%) in \ce{Ar}, Air Liquide Germany)\cite{AirLiquide_Arcal15} and \ce{CO2} (KOHLENDIOXID technisch, Air Liquide Germany),\cite{AirLiquide_CO2} as well as to an in-house supply line providing \ce{N2} or compressed air. Check-valves (CVs 01-03 Swagelok, USA) downstream of the MFCs prevent backward diffusion of gas.

\subsubsection*{(ii) Reaction cell}
The IR tube furnace cell by Bischoff \textit{et al.}\cite{Bischoff2024} is based on previously applied diffraction studies (Andrieux \textit{et al.}\cite{andrieux_high-pressure_2014} and Doran \textit{et al.}\cite{doran_compact_2017}). The sample is placed inside a fused silica quartz glass tube (supplied by CM Scientific, Ireland, or Hilgenberg, Germany). Tubes with varying wall thicknesses were tested in this study, including both open-ended tubes and those with one closed end. Tubes with both ends open enable a plug-flow mode inside the reactor cell. Dimensions of the tubing used for the \textit{in situ} and \textit{operando} studies are 100\,$\pm$\,1\,mm in length, 0.80\,$\pm$\,0.08\,mm inner diameter and 1.0\,$\pm$\,0.1\,mm outer diameter (1.0/0.8-capillary). The capillary walls provide suitably low absorption at the Mn and Ni K-edges and can withstand pressures of at least 11\,bar, as verified in independent tests outside the experimental setup (higher pressures could not be tested due to site limitations). If needed, an optional pressure control unit (PC) can be installed between the reaction cell and Bubbler B-01 to regulate pressure during operation. Other tubes, with dimensions of 100\,$\pm$\,0.5\,mm in length, 1\,$\pm$\,0.1\,mm inner diameter, and 1.5\,$\pm$\,0.1\,mm outer diameter (1.5/1.0-capillary), were tested. These can operate at pressures up to 50\,bar and temperatures of 1000\,°C.\cite{Bischoff2024} However, due to their higher wall thickness, they are only suitable for absorption edge energies above 11\,keV, such as the Pt L-edge at 11\,564\,eV\cite{Elam2002} or the Zr K-edge at 17\,998\,eV\cite{Elam2002} when used in this spectrometer. Fig.~\ref{fig:cell}(a) shows schematically the positioning of the sample inside the glass tube. In the center of the capillary, a 1\,mm-long quartz wool segment is placed. Upstream of this quartz wool segment, about 5\,mm of the sample is placed, followed by a second wool segment to prevent the sample from moving, while mounting the capillary. Downstream of the first quartz wool segment, a Ø0.5\,mm thin K-type thermocouple (TC; Reckmann, Germany) is placed. The symmetrical design positions the thermocouple (TC) tip and the sample at equal distances from the center of the tube, thereby positioning them in "isothermal points".\cite{Bischoff2024} In the experimental setup, the entire capillary system is oriented vertically, with the TC positioned at the bottom, as shown in Fig.~\ref{fig:cell}(c). This configuration helps to prevent sample displacement during measurements due to gas flow and thermal expansion. In addition, placing the TC at the bottom below the catalyst bed ensures that the gas first interacts with the sample, preventing any changes in gas-phase concentration that might otherwise result from contact with the hot K-type thermocouple. Mixed vespel/graphite ferrules (Mascom, Germany) with a Ø1.1\,mm hole are used to enclose the glass capillary and seal it via Swagelok nuts and two T-pieces. These T-pieces are connected to quick-connect fittings, enabling rapid attachment to the setup piping without the need for screwing. Before attaching the second T-piece, a silicon carbide (SiC) tube is assembled around the capillary (see Fig.~\ref{fig:cell}(a) and (b) and Fig.~S2 and S3 in the SI). The SiC tube functions as a furnace, with two 4\,mm long and 1\,mm wide slits cut into opposite sides using a diamond cutting disc, allowing X-rays to pass through the capillary and the sample within. SiC is highly suitable as a furnace material due to its high hardness -- even at elevated temperatures -- as well as its good IR absorption and favorable X-ray absorption properties.\cite{doran_compact_2017,schlicker_transmission_2018}

The filled capillary, together with the SiC tube and T-pieces including the quick-connect fittings, is mounted on a transport plate (see Fig.~S2 and S4 in the SI) to simplify assembly and prevent damage to the tube during transport. The assembled transport plate is subsequently inserted into the main furnace setup (see Fig.~\ref{fig:cell}(a) and Fig.~S2 in the SI), which consists of two IR halogen lamps (64635 HLX 150\,W, Osram, Germany). These lamps heat the SiC tube, whose walls are positioned at the focal point of the lamps (19.5\,mm). The tube is held in place by two U-shaped stainless steel brackets. To prevent rotation, the tip of a screw is gently pressed against a flat indent on the SiC tubing (see Fig.~S4 in the SI).

For more information on the reactor cell, including the performance,  we refer to Bischoff \textit{et al.}\cite{Bischoff2024} The cell can be isolated by closing two valves in the up- and downstream of the cell (see Fig.~\ref{fig:flow-chart}).

\subsubsection*{(iii) Downstream: Gas analysis}
 The downstream section (iii) contains two bubblers (B-01 and B-02) and a GC unit, to measure the catalytic activity, which is further explained in Section~\ref{sec:GC}. The first bubbler (B-01), located upstream of the GC units, is used to trap moisture and prevent it from entering the GC system. The second bubbler (B-02) is filled with water and is used to indicate the gas flow through the cell.


\subsection{\label{sec:GC}Gas chromatography (GC)}
To measure the catalytic activity of the samples (in particular, \ce{CO2} conversion as well as \ce{CH4} and CO production) an online gas chromatograph (GC, Micro GC Fusion 2-Module system, INFICON GmbH, Cologne, Germany)\cite{inficon_micro_gc_fusion} equipped with Rt-Molsieve 5\,\si{\angstrom} (in module A) and Rt-Q-Bond (in module B) columns, utilizing a thermal conductivity detector (TCD) was employed. Helium (He) and argon (Ar) gases were connected to the GC, each regulated to 4\,bar pressure. Both gases served as carrier and reference gases for the two modules, ensuring accurate measurement of thermal conductivity differences between the sample and carrier gas in the TCD.

\subsection{\label{sec:samples-references}Samples \& references}

\subsubsection{\label{sec:5NiMnO}5\,\% Ni/MnO catalyst}
The 5\,\% Ni/MnO catalyst used to investigate the \textit{in situ} oxidation of MnO to \ce{Mn2O3} was synthesized as described by Gili \textit{et al.}\cite{Gili2018, Gili2025-2} Specifically, it was prepared via a coprecipitation method using \ce{Ni(OCOCH3)2.4H2O} (nickel acetate tetrahydrate, Sigma-Aldrich) and \ce{Mn(NO3)2.4H2O} (manganese nitrate tetrahydrate, Sigma-Aldrich) as metal precursors, corresponding to a Ni:Mn atomic ratio of 0.05:0.95. The appropriate amounts of these precursors were dissolved in \SI{100}{\milli\litre} of deionized water (DIW). To initiate coprecipitation, a second solution containing \SI{55}{\milli\mole} of \ce{NaHCO3} (Merck) in \SI{50}{\milli\litre} of water was added dropwise under vigorous stirring. The pH of this solution was adjusted to $\geq$10 using \ce{NaOH} (Merck). After stirring for \SI{8}{\hour}, the resulting precipitate was collected by filtration and washed several times with DIW. The solid was then dried under vacuum at \SI{80}{\celsius} overnight, followed by calcination at \SI{750}{\celsius} for \SI{4}{\hour} with a heating rate of \SI{5}{\celsius\per\minute}.

After calcination of the synthesized catalyst, a part of the material was treated in a 5\,\% \ce{H2} in \ce{N2} atmosphere at 500\,°C for at least 60\,min to reduce \ce{NiO/Mn2O3} to \ce{Ni/MnO}. XRD measurements of the treated and untreated material confirm the successful reduction of \ce{Mn2O3} to \ce{MnO} in the material (see Fig.~S5 in the SI).

\subsubsection{\label{sec:20-Ni-COK-12}20-NiO/COK-12 catalyst}
The synthesis of the COK-12 (ordered mesoporous \ce{SiO2}) support was performed in a batch upscaled by a factor of 50, as previously reported.\cite{Jammer2009} Briefly, 200\,g of Pluronic P123 (MW \SI{5800}{\gram\per\mole}, from Sigma-Aldrich, Merck, Germany) was dissolved in \SI{5375}{\milli\liter} deionized water (DIW). To this solution, 168.1\,g of anhydrous citric acid (\(\geq\)99.5\,\%, from Carl Roth GmbH+Co.KG, Karlsruhe, Germany), 144.1\,g of trisodium citrate dihydrate (\(\geq\)99\,\%, from Carl Roth GmbH+Co.KG, Karlsruhe, Germany), and a solution of 520\,g of sodium silicate (7.8--8.5 wt.\% \ce{Na2O}, 25.8--28.5 wt.\% \ce{SiO2}, from Carl Roth GmbH+Co.KG, Karlsruhe, Germany) in \SI{1500}{\milli\liter} DIW were incorporated. After a day of aging at RT, the precipitated solids were filtered from the aqueous solution, washed with \SI{25}{\liter} DIW, and dried at \SI{60}{\celsius}. The resulting powder was calcined at \SI{500}{\celsius} for two hours to remove the organic template.

For metal loading, incipient wetness impregnation followed by vacuum drying was performed on the COK-12 support. 0.515\,mL/g$_{\textnormal{COK-12}}$ of a nickel nitrate hexahydrate aqueous solution (24.64\,g \ce{Ni(NO3)2 \cdot {6}H2O}, from Merck, Darmstadt, Germany in \SI{10}{\milli\liter} DIW) was added to 1\,g of COK-12. The resulting light-green powder was vacuum-dried immediately after deposition using a vacuum furnace (VT 5042 EK, Heraeus, Germany) at RT for approximately 2 hours until dry. The powder was then placed in contact with a concentrated ammonia solution (\ce{NH3 \cdot H2O}, 25\,\%, from Carl Roth GmbH+Co.KG, Karlsruhe, Germany), causing a change of color to light blue due to the formation of the nickel-ammonia complex, after which it was vacuum-dried a second time. The obtained pre-catalyst, which contained approximately 20.2 wt.\% NiO (corresponding to 15.8 wt.\% Ni)  after calcination as determined by quantitative X-ray fluorescence (XRF, see SI), is referred to as 20-NiO/COK-12 in this paper.

Prior to catalytic testing, the GC was calibrated to detect \ce{H2}, \ce{N2}, \ce{O2}, \ce{CO}, and \ce{CH4} in Module~A, as well as \ce{CO2} and C2–C4 hydrocarbons in Module~B. Water could be qualitatively identified in Module~B when required. During the catalytic experiments \ce{CH4} was the only carbon-containing product detected, with overall carbon mass balances exceeding 93\,\%. \ce{CO} was below the detection limit of the GC. Flow compression effects were neglected due to the low conversion and minimal overall change in total mole number.

The reactions taken into account in the system are methanation and RWGS (\ref{eqn:methanation} and \ref{eqn:rwgs}). The calculations for conversion, and selectivity were performed utilizing the dgbowl package.\cite{Senocrate2024,kraus_dgpost_2024} 

In summary, product- and carbon-based \ce{CO2} conversion ($X_{\ce{CO2}}$), was calculated as:

\begin{equation}
X_{\ce{CO2}} = 100 \cdot \frac{ [\ce{CO}]_o + [\ce{CH4}]_o -[\ce{CO2}]_o }{[\ce{CO}]_o + [\ce{CH4}]_o}
\label{eq:conversion}
\end{equation}

Where the subindex $o$ refers outlet concentrations (calculated from GC data).

Selectivity towards methane ($S_{\ce{CH4}}$) was calculated as:

\begin{equation}
S_{\ce{CH4}} =100 \cdot \frac{[\ce{CH4}]_o}{ [\ce{CO}]_o + [\ce{CH4}]_o}
\label{eq:selectivity}
\end{equation}

\subsubsection{\label{sec:references-Ni-Mn}Mn and Ni references}
 Different manganese oxides (\ce{MnO}, \ce{MnO2} and \ce{Mn2O3}) that had already been used as references in Dimitrakopoulou \textit{et al.}\cite{Dimitrakopoulou2018} were already prepared on adhesive tape and were employed as references for the \textit{in situ} Mn K-edge investigation of the 5\,\% Ni/MnO catalyst.\cite{Dimitrakopoulou2018} Furthermore, the calcined catalyst material (5\,\% NiO/\ce{Mn2O3}) and the reduced catalyst (5\,\% Ni/MnO), which was measured \textit{in situ}, were also prepared as pellets to investigate artifacts introduced by the capillary and to allow a direct comparison of the sample measured under \textit{in situ} conditions to the same material with different oxidation states. This is particularly important, as the presence of Ni in the sample and the synthesis procedure mentioned above (see Section~\ref{sec:5NiMnO}) could influence the Mn species in the sample and thus differ from the pure Mn oxide reference materials.

 For the \textit{operando} investigation of 20-NiO/COK-12, the material itself has been prepared as a pellet sample for \textit{ex situ} measurement as a reference as well as NiO powder (Sigma Aldrich).\cite{sigma_aldrich_nickel-oxide} The detailed procedure for pellet sample preparation is described in Praetz and Schlesiger \textit{et al.}\cite{Praetz-Schlesiger2025} For the expected metallic state (Ni(0)) after reduction a 10\,\unslant{µ}m Ni-foil (Goodfellow, 99.95\,\%)\cite{Goodfellow_NiFoil_LS83374} was included as reference standard. Roughly 1\,mg of 20-NiO/COK-12 was loaded in the capillary, which yielded detectable activity under the experimental conditions, while being of a short length, minimizing temperature inhomogeneity.\cite{Bischoff2024}

\subsubsection{\label{sec:additional-samples}Additional samples}
The following materials and references were measured to evaluate the capabilities and performance of the \textit{in situ}/\textit{operando} setup combined with the spectrometer across different energies and capillary types.
 
\textbf{SnSe} (tin selenide), a leading candidate for thermoelectric materials,\cite{zhao_ultralow_2014,li_orbitally_2015} which play an important role in sustainable energy conversion by efficiently converting waste heat into electricity, was investigated at the Se K-edge at 12\,658\,eV.\cite{Elam2002} The SnSe powder was prepared from a commercial single crystal (2D Semiconductors, BLK-SnSe),\cite{2dsemiconductors_snse} which was grown using the Bridgman technique. The powder was obtained by gently crushing the bulk crystal with a mortar and pestle.

\textbf{\ce{ZrO2}} (zirconium(IV) oxide, Alfa Aesar)\cite{alfa_aesar_zirconia_P19G036} was measured using a capillary with an inner diameter of 1\,$\pm$\,0.1\,mm and an outer diameter of 1.5\,$\pm$\,0.1\,mm, as used in Bischoff \textit{et al.}\cite{Bischoff2024} This material was selected because the Zr K-edge at 17\,998\,eV\cite{Elam2002} lies at the upper energy limit of the spectrometer used in this work. The pure material was diluted with \textit{Hoechst} Wax C in a ratio of 1:2 to achieve the optimal absorption.



\begin{figure*}[t]
\centering
    \includegraphics[width=180mm,trim = 30 60 30 330]{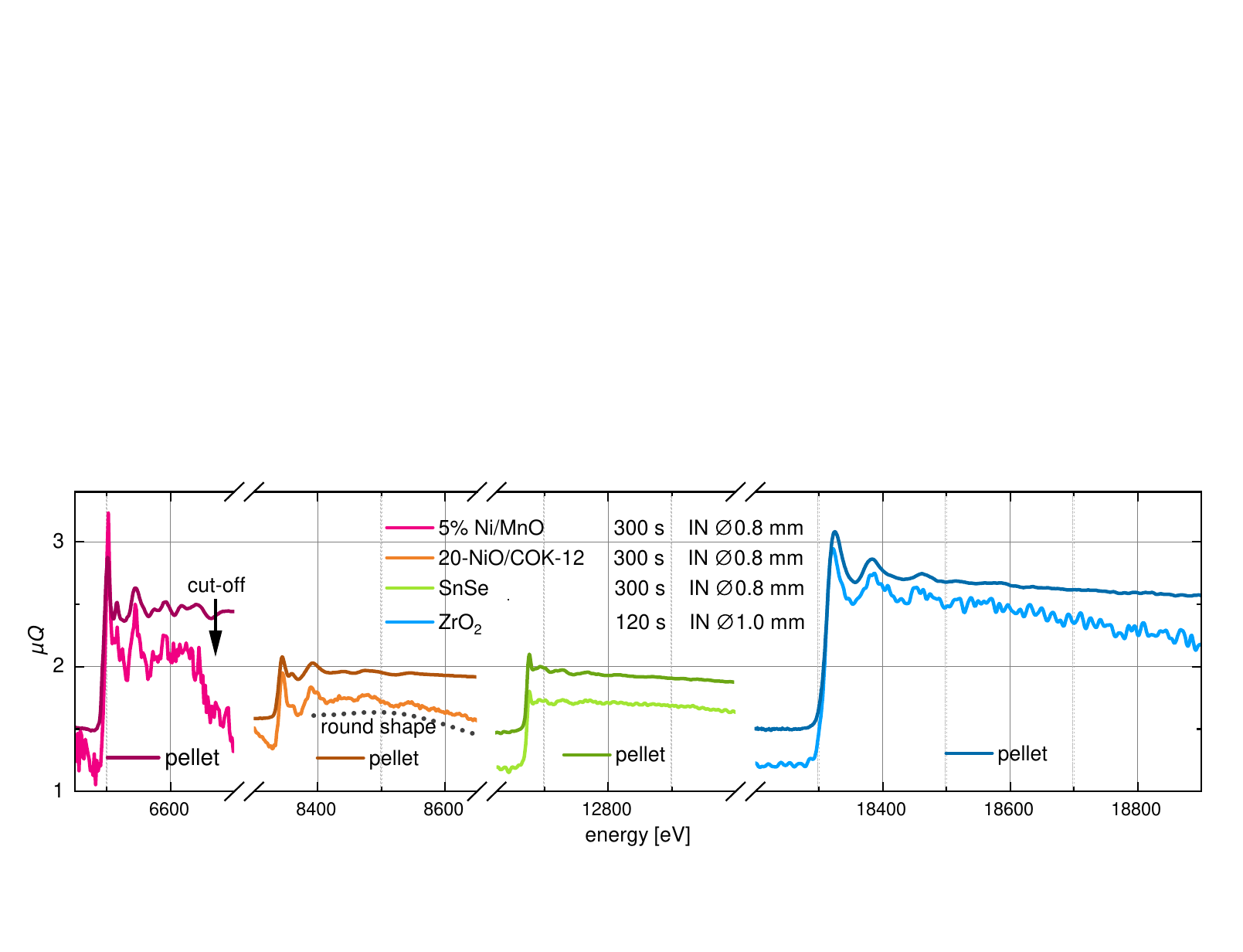}
    \caption{K-edge XAS spectra (Mn, Ni, Se and Zr) of 5\,\% Ni/MnO @ 6\,539\,eV,  20-NiO/COK-12 @ 8\,333\,eV, SnSe @ 12\,658\,eV and \ce{ZrO2} @ 17\,998\,eV. The Mn, Ni and Se K-edge measurements were performed with the capillary with 0.8\,mm inner diameter (IN) and 1.0\,mm outer diameter. The Zr K-edge measurement was performed using the capillary with 1.5 outer diameter and 1.0\,mm inner diameter. Vertical offsets to the overall absorption ($\mu Q$) of the different spectra have been introduced for better comparison with each other. The unmodified spectra are shown in the SI in Fig.~S6--S9. The dark lines on top of the capillary spectrum represent the same materials prepared as a pellet, without distortion introduced by the capillary.}
    \label{fig:overview}
\end{figure*}

\section{\label{sec:results}Results}
This section begins with an overview in Section~\ref{sec:overview}, demonstrating the performance of the setup across different energies and discussing its limitations. Section~\ref{sec:Mn-Oxidation} presents the \textit{in situ} measurements of Mn oxidation in 5\,\% Ni/MnO, followed by the \textit{operando} investigation of the 20-NiO/COK-12 catalyst before, during and after activation in Section~\ref{sec:Ni-Reduktion}. 

In the following, we refer to the first experiment as \textit{in situ}, as it involved the controlled oxidation of the 5\,\% \ce{Ni/MnO} catalyst under a defined temperature and gas environment (compressed air, stepwise heated from RT to \SI{600}{\celsius}). In contrast, the second experiment is designated \textit{operando}, as it was conducted under actual catalytic conditions relevant to \ce{CO2} hydrogenation over a \ce{NiO}/\ce{SiO2} catalyst, including simultaneous gas-phase analysis by gas chromatography. This terminology follows the IUPAC recommendations outlined by Peterson \textit{et al.}\cite{peterson_terms_2024}

The absorption is expressed in units of $\mu Q$, where $\mu$ denotes the mass attenuation coefficient and $Q$ represents the mass deposition of the analyte. Data normalization and further processing were carried out using \textit{ATHENA}, part of the \textit{Demeter} software package.\cite{Ravel2005} Pre-edge subtraction was performed using a linear fit to the available pre-edge region, which remained well defined even for spectra limited to the XANES range by capillary-induced limitation. Post-edge normalization was achieved by fitting a polynomial function (typically second order) to the accessible post-edge region. In spectra affected by a pronounced high-energy cut-off or by rounded spectral shapes arising from the cylindrical capillary geometry, the normalization fit was restricted to the highest-energy interval free of saturation and geometric distortion. To ensure consistency and comparability, identical normalization parameters and fitting ranges were applied to all spectra within a given dataset, using common reference (isosbestic) points where possible. While absolute normalization is constrained by the limited energy window in some cases, this approach reliably captures relative spectral changes, such as oxidation-state evolution during \textit{operando} reduction and reaction experiments.


\subsection{\label{sec:overview}Energy-range performance and capillary-induced effects}
In addition to the 5\,\% Ni/MnO sample discussed in Section~\ref{sec:5NiMnO} and 20-NiO/COK-12 in Section~\ref{sec:20-Ni-COK-12}, several other sample systems and capillary configurations were tested to evaluate optimal conditions for the energy range of interest and overall feasibility. Fig.~\ref{fig:overview} shows four example of K-edge XAS spectra at different energies, spanning nearly the entire energy range currently accessible with the presented spectrometer. For better comparison, vertical offsets have been applied to some of the spectra. The unmodified spectra are provided in the SI (Fig.~S6–S9). The darker lines in Fig.~\ref{fig:overview} represent the same material prepared as a pellet, without the distortion introduced by the capillary. A comparison between the normalized measurements of the material in the loaded capillary and as a free-standing pellet is shown in the SI (Fig.~S10 and S11). The pure materials were diluted when necessary, using boron nitride (BN) for \textit{in situ}/\textit{operando} experiments, or Hoechst Wax C for measurements performed at RT to achieve an absorption of $\mu Q$ around 1. The actual measured absorption and the absorption step itself depend not only on the concentration of the analyte in the material or in the diluted material mixture but also on the stacking behavior of the material/mixture during the capillary filling process, as well as on compression effects caused by inserting a second piece of quartz wool after loading the sample. Consequently, the absorption step observed for the different materials in Fig.~\ref{fig:overview} reflects not only the mass of the analyte per area/volume prior to filling and any potential dilution, but also these packing and mechanical influences. Additionally, during \textit{in situ} and \textit{operando} measurements, this absorption step -- as well as the overall absorption -- can change due to thermal heating, particle movement, and other factors, and therefore must be accounted for.

The first spectrum, 5\,\% Ni/MnO (see Fig.~\ref{fig:overview}, magenta) measured at the Mn K-edge within 300\,s (5\,min) measurement time using the 1.0/0.8-capillary, shows mainly the XANES region with a sharp cut-off of the spectrum at around 6\,650\,eV. This sharp cut-off is due to the limitation by the capillaries, as described in Section~\ref{sec:von-Hamos}. Although the SNR is low, the short measurement is still sufficient to observe changes in the oxidation state, as will be demonstrated in Section~\ref{sec:Mn-Oxidation}.

Secondly, the shown Ni K-edge spectrum of 20-NiO/COK-12 (see Fig.~\ref{fig:overview}, orange) also measured with the 1.0/0.8-capillary within 300\,s of measurement time. The better SNR compared to 5\,\% Ni/MnO is mostly due to the higher emission of the X-ray source and the lower absorption by the capillary walls towards this higher energy. In this case the cut-off of the spectrum is observed at higher energies from the edge position $E_0$ compared to 5\,\% Ni/MnO, which is due to the higher selected energy range and therefore an increased energy range is passing through the capillary at the same time, as described in Section~\ref{sec:von-Hamos}. Distortion in the spectra, such as the round shape (see dotted line in Fig.~\ref{fig:overview}) instead of a steady decreasing absorption, is still observed, indicating a nonhomogeneous sample distribution in the energy band pass, caused by the round shape of the capillary.

The third spectrum shown in Fig.~\ref{fig:overview} (green) corresponds to SnSe, measured at the Se K-edge using the 1.0/0.8-capillary with a measurement time of 300\,s. Distortions due to the round geometry of the capillary are even less pronounced than those observed at the Mn and Ni K-edges. In this energy range, the use of the 1.5/1.0-capillary may also be feasible, potentially achieving sufficient SNR within a practical measurement time of 5–15 minutes -- particularly if operation at elevated pressures (11–50\,bar) is required.

Lastly, \ce{ZrO2} diluted with Hoechst Wax C was measured in a 1.5/1.0-capillary (shown in blue in Fig.~\ref{fig:overview}) at the Zr K-edge. The spectrum shows almost no distortion by the round geometry of the capillary and no observable cut-off. Even with a measurement time of only 2 minutes, the absorption edge and white-line features exhibit a sufficient SNR to investigate oxidation state changes. This result demonstrates, that for higher energies, even the 1.5/1.0-capillary is suitable and can be used for \textit{in situ}/\textit{operando} experiments up to 50\,bar. As a next step, the implementation of a tungsten (W) X-ray source optimized for 30\,keV is currently under investigation. While the source used in this study is limited to a maximum voltage of 20\,keV, the alternative source is expected to provide a significantly higher photon flux in this energy range, thereby enabling shorter measurement times for the same SNR. This would also allow reliable measurement of the EXAFS region with sufficient SNR for Fourier transform analysis and extended structural interpretation.
Preliminary tests of the new X-ray source are encouraging, showing an approximate 3.7-fold increase in photon flux at the Zr K-edge, see Fig.~S28 in the SI. 

\begin{figure*}[t]
 \centering
 \includegraphics[width=105mm,trim = 65 25 80 20,clip]{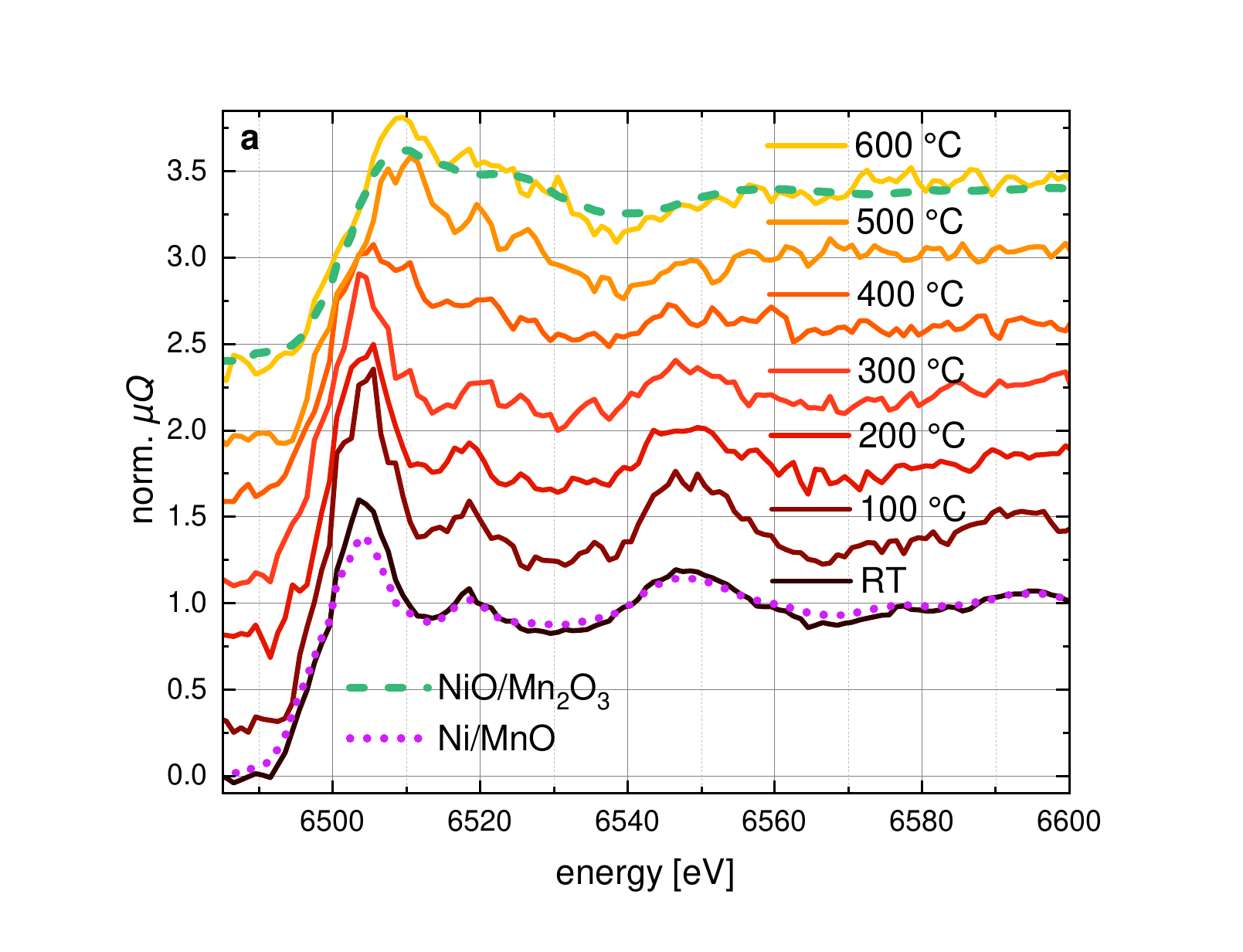}
  \includegraphics[width=79mm,trim = 50 25 250 10, clip]{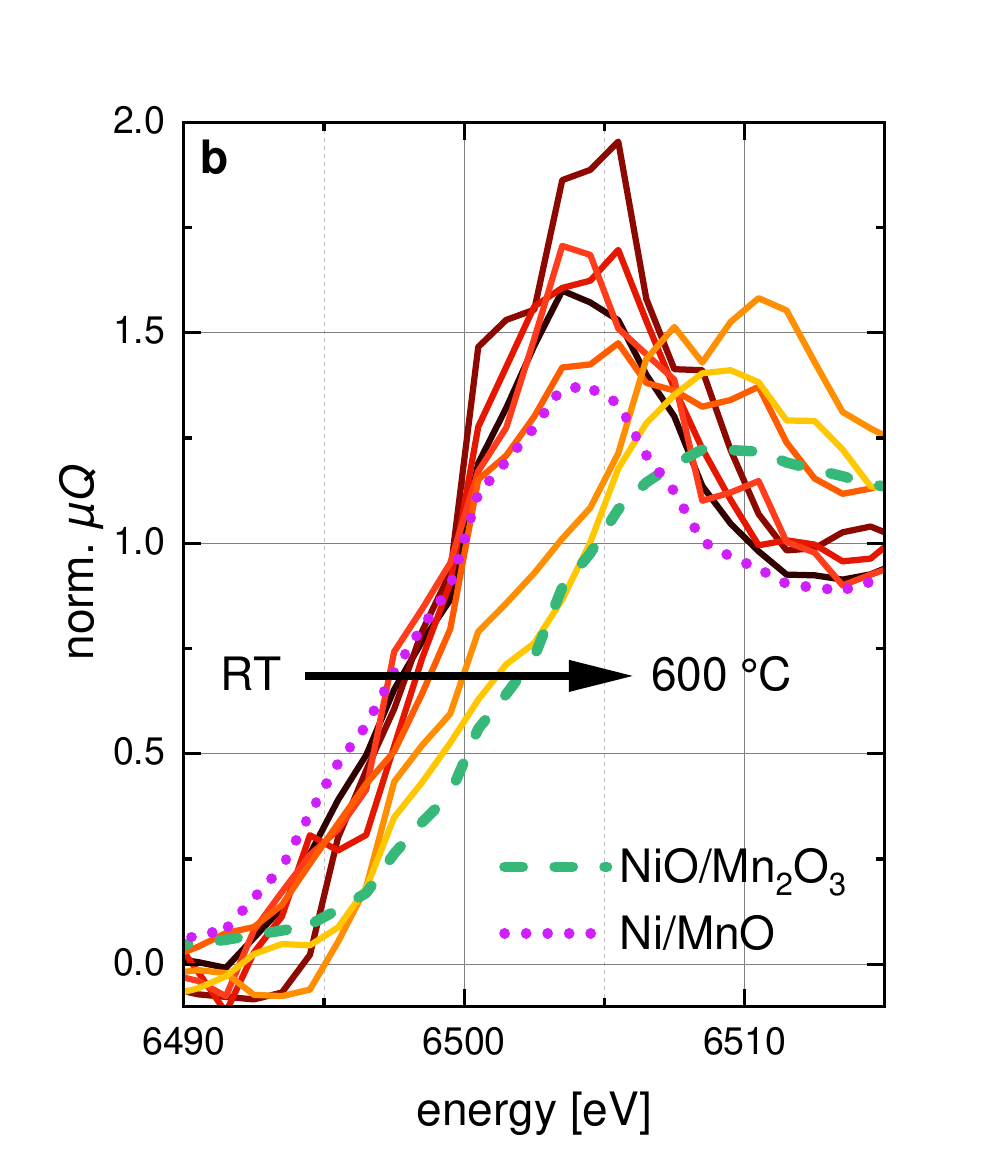}
 \caption{\textit{In situ} XAS measurement of 5\,\% Ni/\ce{MnO} at the Mn K-edge during heating, with a measurement time of 15\,min per spectrum. \textbf{a}: XANES region at RT and elevated temperatures, shown alongside reference spectra of Ni/\ce{MnO} and NiO/\ce{Mn2O3}. An offset has been applied for clarity. \textbf{b}: Enlarged view of the absorption edge, highlighting the Mn K-edge shift toward higher oxidation states upon heating.}
 \label{fig::MnOXidation}
\end{figure*}


\subsection{\label{sec:Mn-Oxidation}\textit{In situ} Mn oxidation in NiO catalyst}
The reduced state of the material (\ce{Ni/MnO}) was used for the \textit{in situ} investigation of the Mn oxidation process by heating \ce{Ni/MnO} from RT to 600\,°C in air in 100\,°C steps and was compared to the calcined and reduced materials measured \textit{ex situ} prepared as free standing pellets. The material loaded in the capillary was diluted with BN in a 1:8 ratio. The measurement time per spectrum was 300\,s. To increase the SNR, the data was 3 x binned, resulting in 15\,min measurement time per spectrum per temperature step.

Fig.~\ref{fig::MnOXidation} shows the results from Mn K-edge \textit{in situ} XAS measurement compared to the reduced (5\,\% Ni/MnO) and oxidized (NiO/\ce{Mn2O3}) material during compressed air flow of about 3\,ml/min at normal temperature and pressure (NTP, 25\,°C and 1\,bar). The reduced flow rate was attributed to partial clogging of the capillary by the packed powder sample, likely caused by the packed powder sample. Significant spectral changes, and thus changes in the Mn oxidation state, began around 400\,°C, with nearly complete oxidation of MnO to \ce{Mn2O3} observed approximately 15\,minutes after reaching 600\,°C. Repeated measurements with different dilutions of the raw materials yielded consistent results.

While a shift in the edge position is observable (see Fig.~\ref{fig::MnOXidation}(a)), accurately determining the exact edge position is challenging due to the low signal-to-noise ratio (SNR), which introduces significant uncertainty. To estimate the relative contributions of MnO and \ce{Mn2O3} at different temperature steps, linear combination fitting (LCF) was performed using \textit{ATHENA}.\cite{Ravel2005} It is important to note, that the values obtained from LCF represent the fractional contributions of each reference spectrum -- \textit{i.e.}, the chemical species of Mn -- to the overall XAS spectrum. These fractions do not directly correspond to weight or atomic percentages in the bulk sample. To determine the actual weight percent of each species, the fractions obtained from LCF should be combined with the known oxide content and the corresponding molar masses. The reference spectra used for the fitting were Ni/\ce{MnO} and NiO/\ce{Mn2O3}, both prepared as pellets. 

The LCF results for the various temperature steps are presented in Fig.~S12--S14, with a corresponding plot of the component fraction as a function of temperature shown in Fig.~S15 in the SI. In the component plot (Fig.~S15, SI), a partial oxidation of MnO towards \ce{Mn2O3} is first observed at 200\,°C. At 400\,°C, the oxidation progresses further, with LCF indicating a \ce{NiO/Mn2O3} component fraction of 47~$\pm$~4\,\%. This suggests, that nearly half of the Mn present in the sample exists in the +III oxidation state at this point. The progression of oxidation is not only supported by the LCF results, but is also clearly reflected in the normalized XANES spectra shown in Fig.~\ref{fig::MnOXidation}(b), which display a distinct shift of the absorption edge toward higher energies -- a spectral signature characteristic of the Mn(II) to Mn(III) transition. At 600\,°C, the oxidation appears nearly complete, with 97~$\pm$~5\,\% \ce{NiO/Mn2O3} component fraction identified by LCF. All LCF results including R-factor and Reduced chi-square ($\chi^2_\nu$) are listed in Table~S1 in the SI.


\subsection{\label{sec:Ni-Reduktion}\textit{Operando} Ni reduction in 20-NiO/COK-12 catalyst for \ce{CO2} methanation}

The calcined catalyst 20-NiO/COK-12 was investigated in three steps. First, as is, under reaction conditions, with \ce{H2} and \ce{CO2} gas flow in the ratio of 4:1 (19.2\,mL/min 5\,\% \ce{H2}/\ce{Ar}, 0.6\,mL/min \ce{CO2}, at NTP) at 350\,°C, without a catalytic activation step. Second, during activation of the catalyst with a \ce{H2} gas flow of 14.6\,mL/min (at NTP) at temperature of 600\,°C. Third, under reaction conditions with \ce{H2}/\ce{CO2} flow at 4:1 ratio (19.2\,mL/min 5\,\% \ce{H2}/\ce{Ar}, 0.6\,mL/min \ce{CO2}, NTP) at 350\,°C after activation and reduction of Ni in the catalyst. 

The MFC for \ce{CO2} was operated near its lower control limit, likely causing deviations from the nominal flow. Based on the nominal MFC settings, the \ce{H2}/\ce{CO2} ratio would be about 1.6:1 instead of the intended 4:1, and based on the 5\,\% \ce{H2} flow rate, the \ce{CO2} flow is probably around 0.23\,mL/min. However, GC analysis (see Fig.~S27(c) in the SI) confirmed that the effective ratio in the reaction stream was close to 4:1. Therefore, the gas flows were adjusted accordingly, and the reported \ce{CO2} flow should be considered nominal rather than exact. 

\begin{figure}[t!]
\centering
 \includegraphics[width=80mm,trim = 90 0 80 85]{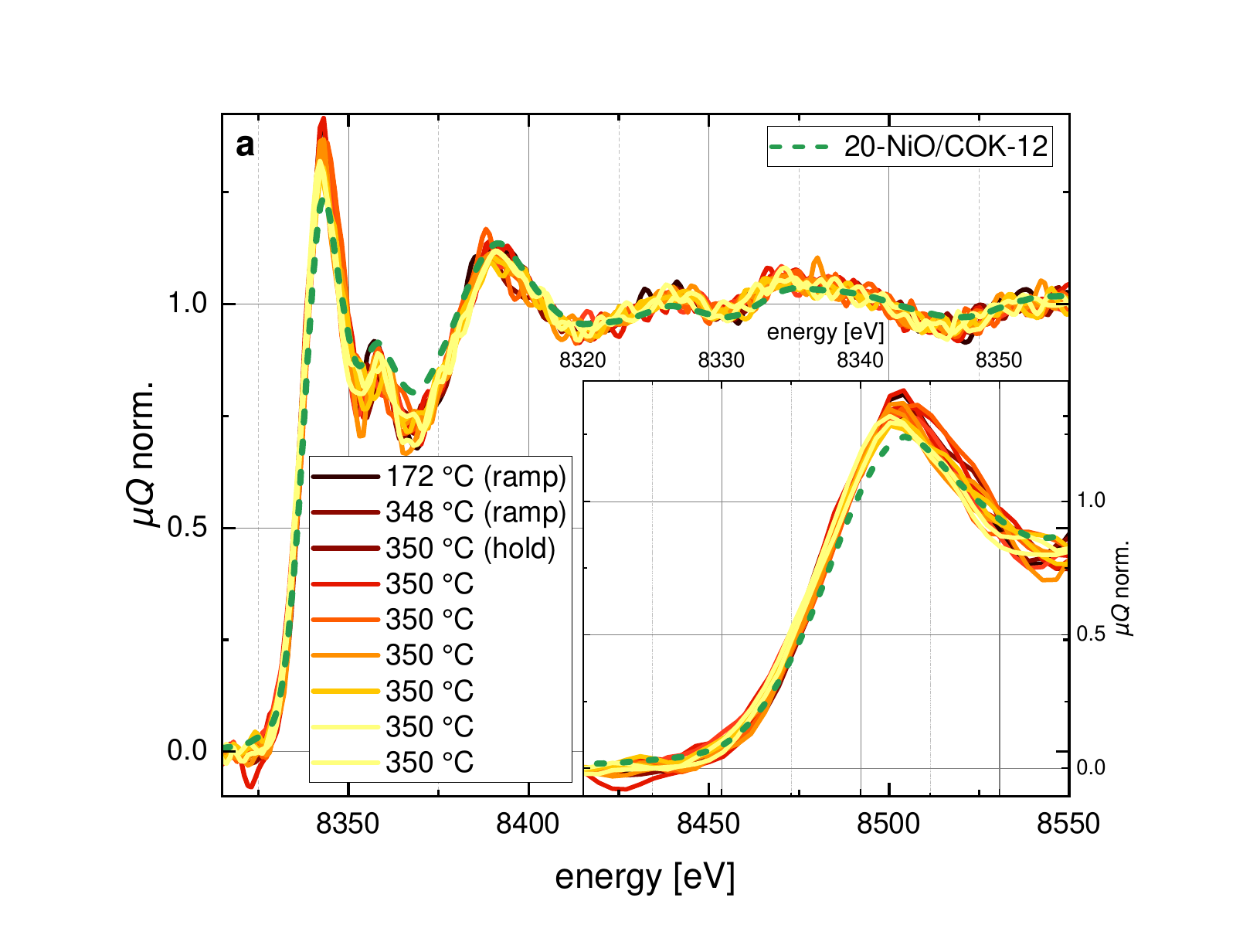}
  \includegraphics[width=80mm,trim = 90 0 80 85]{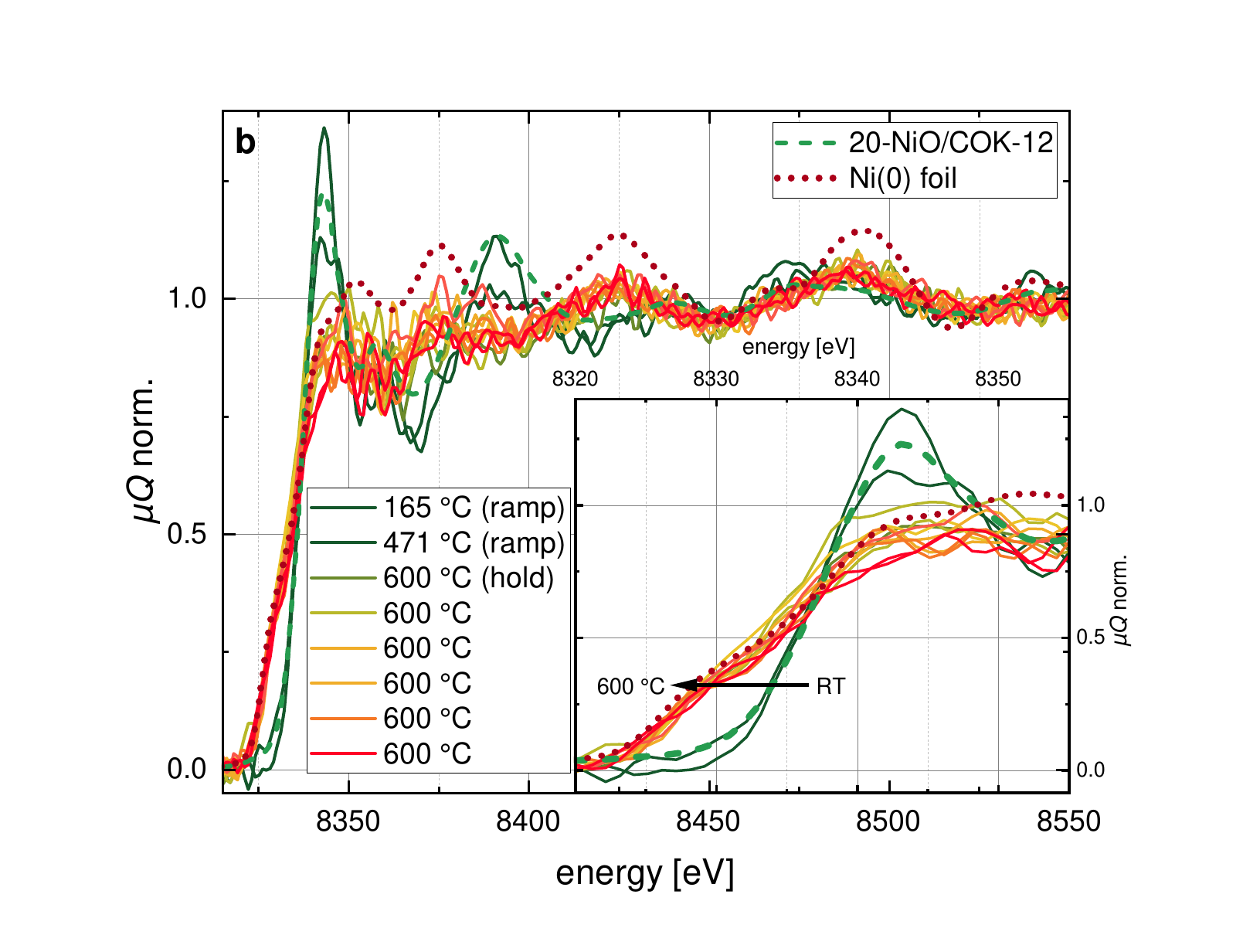}
 \includegraphics[width=80mm,trim = 90 25 80 85]{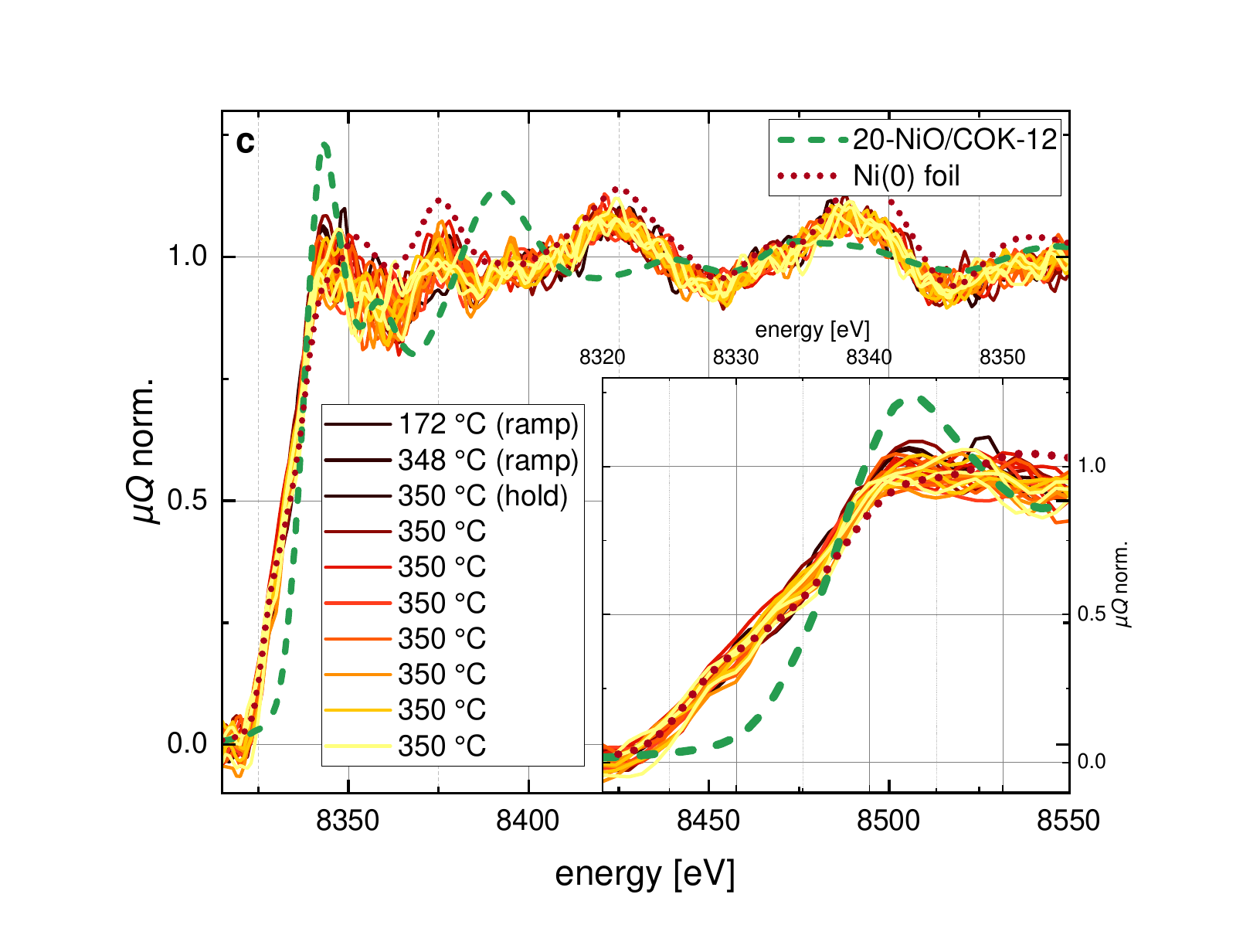}
    \caption{\textit{Operando} XAS measurements of 20-NiO/COK-12 at the Ni K-edge. \textbf{a:} Before catalyst activation, measured at 350\,°C for 1\,h under \ce{H2}/\ce{CO2} flow at a 4:1 ratio. \textbf{b:} Catalyst reduction/activation at 600\,°C for 1\,h under 5\,\% \ce{H2}/Ar flow. \textbf{c:} After activation, measured again at 350\,°C for 2\,h under the same \ce{H2}/\ce{CO2} (4:1) conditions as the initial state. Each spectrum was acquired with a measurement time of 300\,s (5\,min). The results are shown alongside the reference spectra of 20-Ni0/COK-12 prepared as pellet and a 10 µm Ni(0) metal foil with a total acquisition time of 3.5\,h and 1.5\,h, respectively.}
  \label{fig:Operando-NiO-Ni}
\end{figure}

\begin{figure*}[b]
 \centering
 \includegraphics[width=180mm,trim = 8 95 15 320,clip]{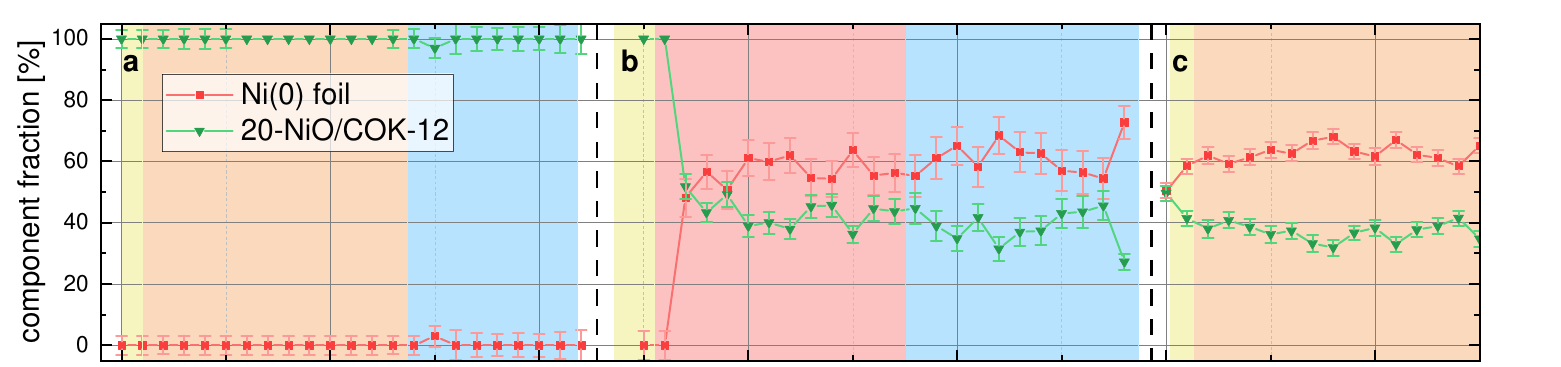}
 \includegraphics[width=180mm,trim = 8 95 15 320,clip]{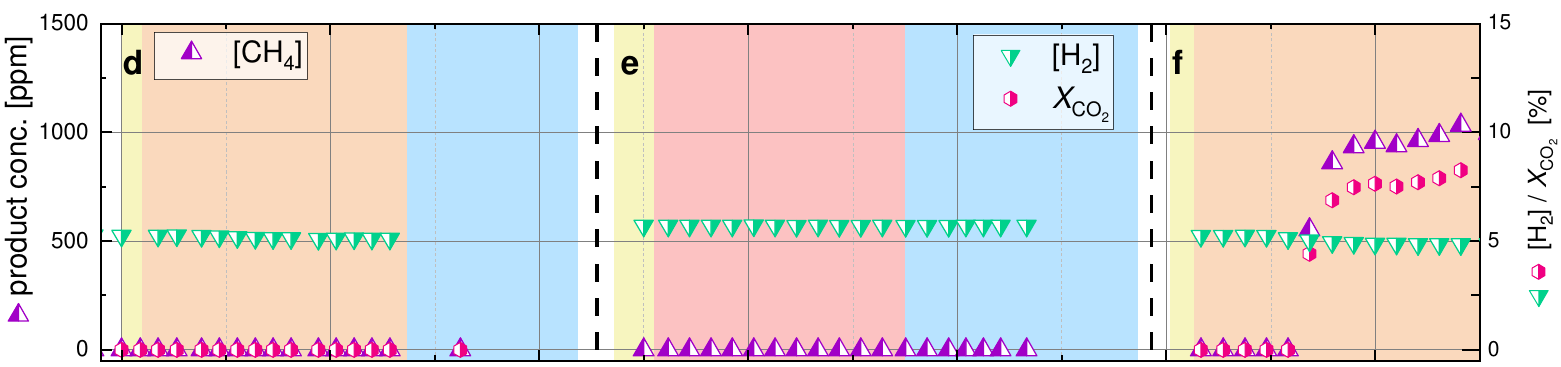}
 \includegraphics[width=180mm,trim = 8 60 15 320,clip]{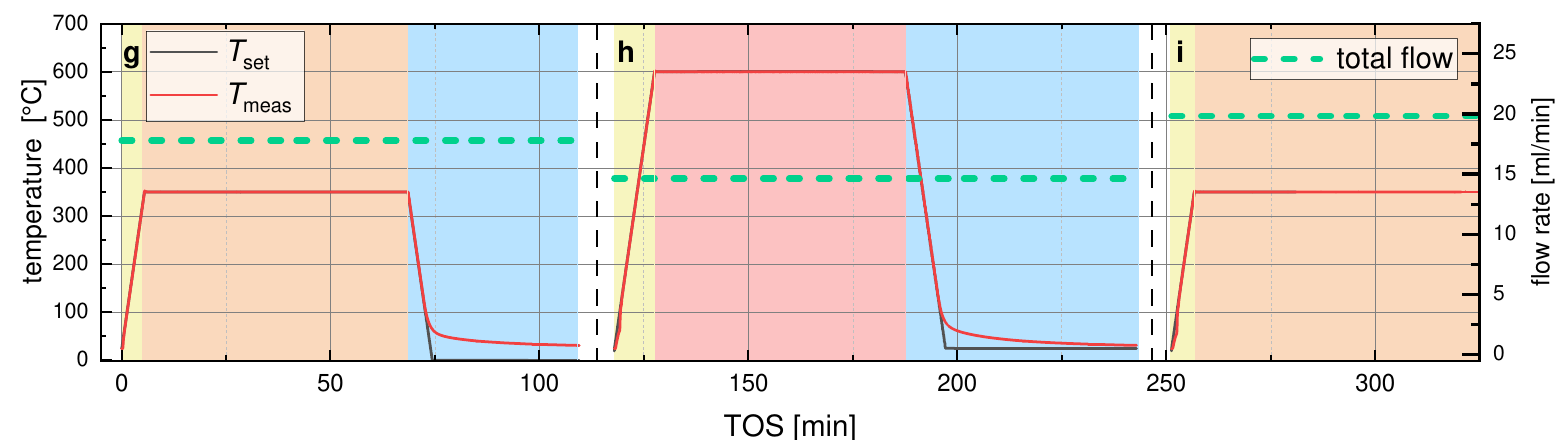}
 
 \caption{\textit{Operando} measurements of the 20-NiO/COK-12 catalyst. The top panels (a–c) show the component fractions obtained by LCF of the XAS spectra, the middle panels (d–f) the corresponding GC analysis, and the bottom panels (g–i) the applied temperature profile and total flow rate. All data are displayed on a common cumulative time axis (time-of-stream, TOS) for clarity, with dashed lines indicating the boundaries between the three experimental steps: catalytic conditions before activation/reduction (a, d, g), during activation/reduction (b, e, h), and after activation/reduction under catalytic testing (c, f, i) with a WHSV of \SI{1.2e6}{\milli\liter\per(\hour\cdot\gram_{cat})}. Note that the time axis represents concatenated experimental stages and does not correspond to the actual elapsed chronological time. $X_{\ce{CO2}}$: \ce{CO2} conversion; [\ce{H2}]: \ce{H2} concentration; [\ce{CH4}]: \ce{CH4} product concentration.}

 \label{fig::Pre-Red-After-3x3plot}
\end{figure*}

Additional supplementary experiments at higher temperatures (400\,°C), as well as doubling the gas flow at 350\,°C, were conducted to further investigate the performance of the activated catalyst. The temperature ramp for all three steps was 1\,°C/s. The sample material was not diluted before being inserted into the capillary. 
The \textit{operando} XAS results at the Ni K-edge for the three measurement steps are presented in Fig.~\ref{fig:Operando-NiO-Ni}. LCF, as described in Section~\ref{sec:5NiMnO}, was performed using Ni(0) foil and 20-NiO/COK-12 (prepared as a pellet) as reference materials. The resulting component fractions are shown in Fig.~\ref{fig::Pre-Red-After-3x3plot} for each step (a–c), alongside the GC analysis (d–f) and the corresponding temperature and flow rate (g–i).  

Fig.~\ref{fig:Operando-NiO-Ni}(a) shows the 20-NiO/COK-12 catalyst under catalytic conditions (as previously described) prior to activation. No significant changes are observed in the normalized \textit{operando} XAS spectra. Additionally, GC analysis of the outlet gas reveals no detectable formation of CO or \ce{CH4} (see Fig.~\ref{fig::Pre-Red-After-3x3plot}(d) and Fig.~S25 in the SI), indicating the absence of catalytic activity at this stage, as expected. However, differences compared to the 20-NiO/COK-12 reference spectrum (measured as a pellet, shown as a dashed line) are noticeable, particularly at the white line (\textit{i.e.}, the first peak beyond the absorption edge). These deviations are likely attributed to the sample geometry within the capillary. Alignment of the capillary/cell-system was performed by choosing the position with the maximum edge jump, which may lead to an increased absorption at the edge and, consequently, at the white line, compared to the rest of the spectrum. Additionally, the normalization process of the \textit{operando} XAS spectra is more challenging compared to regularly prepared samples (\textit{e.g.}, pellets) due to spectral distortions caused by the capillary geometry (see Fig.~S16, S17, and S19 in the SI). This effect may be one possible factor contributing to the observed differences in the white line intensities. The spectra of the 20-NiO/COK-12 pellet appear slightly damped over the entire energy range compared to the \textit{operando} measurements, which could be caused by pinhole effects within the pellet itself. Moreover, the unnormalized absorption edge step of the 20-NiO/COK-12 pellet sample is significantly higher ($\sim$1.3) compared to that of the capillary measurement ($\sim$0.5). A high absorption step can lead to damping in the XANES region, as shown by Praetz and Johansen et al.\cite{Praetz-Johanson2025}, and may therefore also explain the differences in the white line. The unnormalized spectra are presented in Fig.~S16 in the SI, additionally covering the application of the gas flow prior to heating, and the cooling process back to room temperature after heating. 

Upon introduction of the \ce{H2} gas flow (Fig.~S16, top left), a general increase in intensity is observed, particularly in the EXAFS region beyond the absorption edge. This increase stabilizes after approximately 10 to 15 minutes and may be attributed to slight particle rearrangement or compaction of the powdered sample within the capillary. After subtracting the contribution from the capillary itself, the effective mass loading is found to increase by approximately 10\,\%, indicating a corresponding rise in the packing density of the sample. During heating to 350\,°C, no changes in spectral features or overall absorption are observed -- as discussed above -- see Fig.~S16, top right in the SI. During the subsequent cooling to RT, a pronounced shift in overall absorption occurs after approximately 10 minutes (Fig.~S16, bottom left), which may be caused by thermal stress on the capillary. This could lead to slight displacement of the capillary or further particle movement. Further packing or settling of the sample may also have contributed to the increased absorption. However, the overall spectral features appear to remain unaffected, which is also shown by a comparison of the first spectrum during flow and a longer measurement after the treatment, see Fig~S16, bottom right. 

In Fig.~\ref{fig:Operando-NiO-Ni}(b), the 20-NiO/COK-12 catalyst is shown during the reduction/activation procedure under 5\,\% \ce{H2}/Ar at a flow rate of 14.6\,ml/min (NTP). The corresponding unnormalized spectra (Fig.~S17 in the SI) cover the sequence of gas introduction, heating, and subsequent cooling. A temperature of 600\,\textdegree{}C is reached within 10 minutes. Spectral changes become evident at an average temperature of 471\,$\pm$\,85\,\textdegree{}C (the large uncertainty arising from the ongoing temperature ramp), corresponding to the second 5-minute spectrum acquired during heating, which is also shown by the LCF results in Fig.~\ref{fig::Pre-Red-After-3x3plot}(b). No drop in the \ce{H2} concentration is appreciable, due to the small amount of catalyst being reduced.

Fifteen minutes after the onset of heating (\textit{i.e.}, during the ``600\,\textdegree{}C hold'' phase in Fig.~\ref{fig:Operando-NiO-Ni}(b)), the sample remained at 600.0\,$\pm$\,0.1\,\textdegree{}C for five minutes. During this period, a complete shift of the absorption edge toward the Ni metal reference (see inset) is observed. In addition, the intensity of the white line is markedly reduced, indicating a significant reduction of NiO to metallic Ni(0). After 65\,minutes, no further spectral changes are qualitatively observed, suggesting that the reduction has reached completion, and the cooling process to room temperature was initiated.

\begin{figure}[t]
\centering
 \includegraphics[width=85mm,trim = 90 15 80 60]{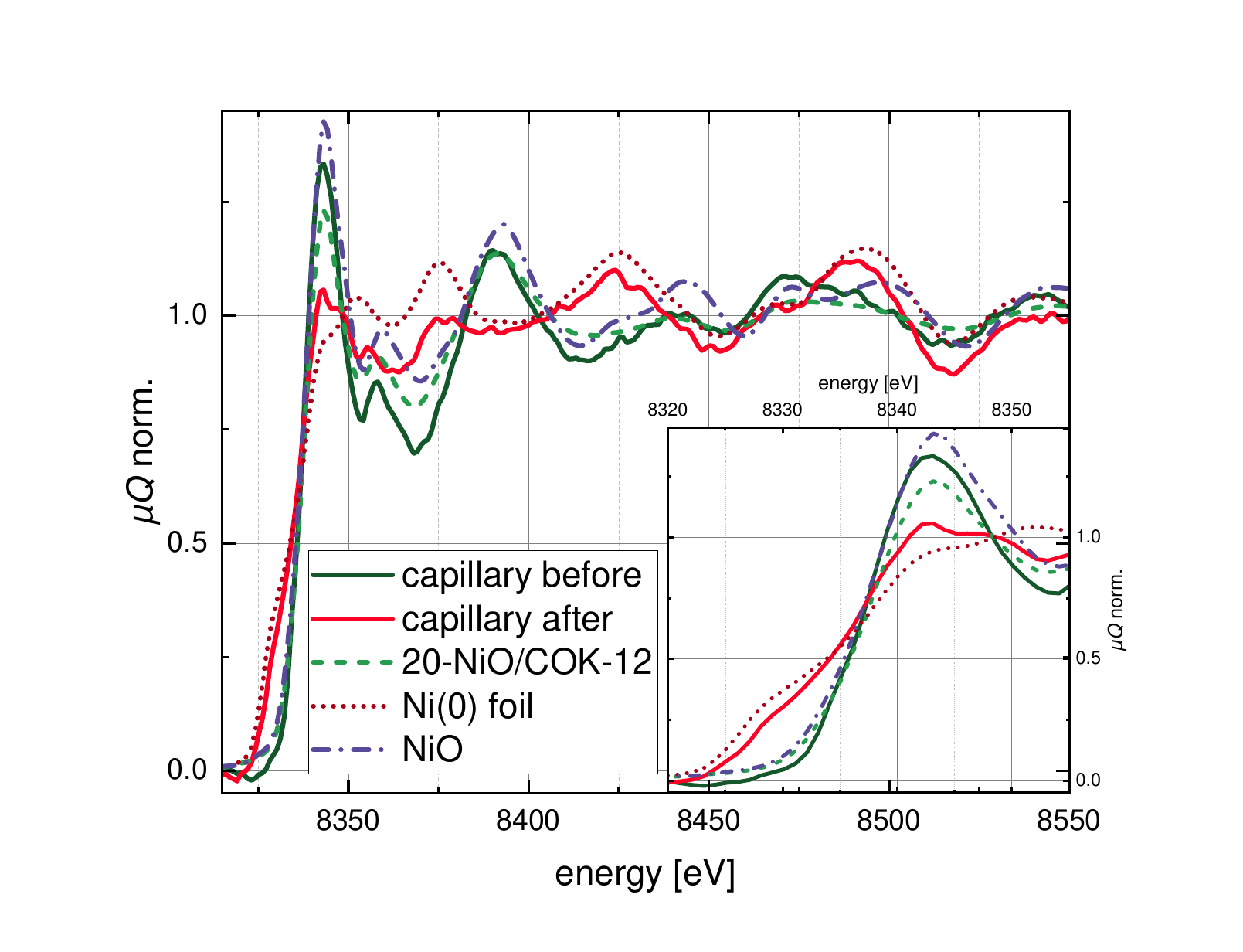}
    \caption{Normalized XAS measurement of 20-NiO/COK-12 before and after reduction/activation procedure, with 50\,min measurement time in comparison to 20-NiO/COK-12 prepared as free standing pellet, NiO and a 10\,µm Ni foil.\cite{Goodfellow_NiFoil_LS83374}}
  \label{fig:Before-after-norm}
\end{figure}

During the cooling phase, oscillations in the EXAFS region of the spectrum (8400--8600\,eV) become more pronounced as the Debye-Waller damping effect -- prominent at elevated temperatures -- is reduced. A 50-minute long measurement taken after cooling the cell to RT shows strong agreement between the treated catalyst and metallic Ni(0), particularly in the EXAFS region (see Fig.~\ref{fig:Before-after-norm}). However, differences in the XANES region indicate that the reduction of the material (Ni) is not fully complete. This suggests the presence of remaining particles in a \ce{Ni^{2+}} state, which can be interpreted as core-shell Ni/NiO particles. Deviations of the reduced sample from that of the pure Ni(0) foil reference may also arise due to interactions of the Ni with the support material (COK-12), also known as metal-support interaction (MSI).\cite{Wang2024} This would also explain the discrepancy of the catalyst sample prior reduction from the NiO reference spectra (see Fig.~\ref{fig:Before-after-norm}) which may also be related to the influence of the support (COK-12, silica) on the electronic structure of the Ni nano particles. This effect is known to stabilize Ni nano particles in silica matrices, sometimes also forming Ni silicates.\cite{Ning2023,Xie2023, Wagner1979} This interpretation is supported by the slightly higher edge position for 20-NiO/COK-12 compared to NiO, see Table~S2 in the SI. The differences between the NiO reference standard and 20-NiO/COK-12 prior to reduction are illustrated in Fig.~S18 in the SI, which presents the normalized spectra of pellet samples along with the residuals, highlighting deviations in the EXAFS region. 

To evaluate the Ni reduction state, LCF was performed using different fitting ranges -- the XANES region, the EXAFS region, and the full spectrum -- in order to assess the robustness of the results. The component fractions shown in Fig.~\ref{fig::Pre-Red-After-3x3plot}(b) were obtained from LCF over the complete XAS range (from 20\,eV below the edge to 250\,eV above it), while the additional fitting ranges were only applied to the reduction measurements for comparison. Fitting in the XANES region primarily probes changes in oxidation state,\cite{Kelly2008} whereas the EXAFS region provides information about the local structural environment, such as the type and distance of neighboring atoms. Applying these different approaches, LCF of the second-to-last 5-minute spectrum after cooling -- while the final spectrum appeared as an outlier and was therefore not considered -- yielded a Ni(0) fraction of 55--72\,\%, whereas the 50-minute post-cooling spectrum indicated a metallic Ni(O) fraction of 62--74\,\%. These results consistently suggest that at least 55\,\% of the Ni in the sample was reduced. LCF fits of the last spectrum at 600\,°C across the different ranges, as well as the fractional component plots for the full operando experiment, are provided in Fig.~S20 and S21 of the SI.

To complement the operando series and to evaluate whether extended heating promotes further reduction, preliminary \textit{in situ} XAS measurements (without GC analysis) were performed on 20-NiO/COK-12. The material was first heated to 500\,°C and held at that temperature for approximately 2\,h. After no significant spectral changes were qualitatively observed, the temperature was increased to 550\,°C and subsequently to 600\,°C, both of which induced further changes in the spectra, indicating continued reduction of the material. At 600\,°C, no additional changes were observed after about 1\,h. The experiment was stopped, and the sample was cooled to RT after 100 minutes at 600\,°C. The raw and normalized data are presented in Fig.~S22 and S23 in the SI.

Building on these preliminary observations, LCF of the \textit{in situ} XAS spectra was carried out to quantify the metallic Ni fraction and to track the progression of reduction as a function of temperature and time. The results indicate that reduction continues, albeit slowly, even after 60 minutes at 500\,°C. Increasing the temperature to 550\,°C and 600\,°C markedly accelerates the reduction rate. While the final spectrum at 600\,°C yields a metallic Ni fraction of 66~$\pm$~4\,\%, cooling the sample to room temperature results in an apparent increase to 91~$\pm$~7\,\%. In both cases, a wide fitting range was applied (20\,eV below to 250\,eV above the edge). The observed difference between high-temperature and room-temperature LCF results is most likely not due to additional reduction during cooling, but rather to thermal damping of EXAFS oscillations at elevated temperatures caused by the Debye–Waller effect.\cite{Beni1976, Lee1981} This damping reduces the apparent metallic contribution when compared with reference spectra recorded at room temperature, thereby biasing the high-temperature LCF fits toward lower metallic Ni fractions. Furthermore, high temperature can also affect the XANES region, as shown by Manuel \textit{et al.},\cite{Manuel2012} which may introduce additional uncertainties if the fit is restricted to this region alone. To mitigate such effects, reference spectra should be measured at elevated temperatures (600\,°C in this work) under an inert atmosphere, for example using the heating cell developed by Praetz and Grötzsch \textit{et al.},\cite{Praetz-Groetzsch2025} which enables heating of pellet samples up to and beyond 500\,°C. It should be emphasized that such damping effects are not limited to laboratory measurements but are also commonly observed in synchrotron-based experiments, highlighting the general importance of using temperature-matched reference spectra.

The overall higher component fraction of the Ni metal phase for the \textit{in situ} experiment compared to the \textit{operando} measurement is most likely due to the longer heating period (60 vs 100 min at 600\,°C) and suggest that prolonged heating can enhance the extent of Ni reduction compared to the \textit{operando} series.

The third step of the \textit{operando} measurement -- reaction conditions after the reduction/activation of 20-NiO/COK-12 -- is shown in Fig.~\ref{fig:Operando-NiO-Ni}(c), which presents the XAS measurement, and in Fig.~\ref{fig::Pre-Red-After-3x3plot}(c,f,i), which displays the component fractions from LCF, the GC analysis, and the cell conditions. The unnormalized spectra, covering the application of the gas flow prior to heating and the subsequent cooling process back to room temperature, are provided in Fig.~S19 of the SI. Under reaction conditions with a \ce{H2}/\ce{CO2} flow ratio of 4:1 (19.2\,mL/min 5\,\%\ce{H2}/Ar, 0.6\,mL/min \ce{CO2}, NTP) at 350\,°C, methane (\ce{CH4}) formation was detected by GC analysis, see Fig.~\ref{fig::Pre-Red-After-3x3plot}(e) and S26--S27 in the SI. Around 1000\,ppm of \ce{CH4} were quantified, leading to a product-based (\ce{CO2}) conversion in the order of 10~$\pm$~4\,\%, and a selectivity towards \ce{CH4} of 100\,\%, as expected due to the low experimental pressure. While these results are somewhat different to previously reported catalysts,\cite{Medina2025} the testing conditions (in particular, the weight hourly space velocity (WHSV, here \SI{1.2e6}{\milli\liter\per(\hour\cdot\gram_{cat})}) and the low reactant concentration) are significantly different, and can be a reason for the observed results. 

As expected, no significant spectral changes -- such as an edge shift or modifications in the EXAFS region -- were observed. Minor variations in the spectra, visible in Fig.~\ref{fig:Operando-NiO-Ni}(c) and Fig.~S19 (top right), are most likely attributable to the Debye–Waller effect and particle movement induced by the reaction and gas evolution within the capillary. 

Doubling the \ce{H2} and \ce{CO2} flow rate (see Fig.~S27 in the SI), which occurred around 330\,min after the start of the measurements, led to a decrease in the detected \ce{CH4} concentrations (Fig.~S27(b)). This decrease is attributed to dilution: the higher volume fraction of \ce{H2} and \ce{CO2} in the outlet gas reduced the relative concentrations of \ce{CH4}. Upon increasing the temperature to 400\,°C under the same doubled flow rate, the concentrations of \ce{CH4} increased again, demonstrating the enhanced catalytic activity at elevated temperature. In contrast, raising the temperature from 300\,°C to 400\,°C did not produce any detectable spectral changes compared to the initial heating from RT to 300\,°C (see Fig.~S19, bottom left in the SI).




\section*{Conclusions}

Using a laboratory von Hámos spectrometer with a plug-flow fixed-bed cell reactor,\cite{Bischoff2024} \textit{in situ} and \textit{operando} XAS measurements were successfully conducted.

\textit{In situ} XAS measurements at the Mn K-edge on the 5\,\% Ni/MnO catalyst were performed to investigate the oxidation of MnO to \ce{Mn2O3}, which initiated at approximately 400\,°C and was tracked using 15-minute spectra. Although edge position determination was limited by the low SNR and thus only qualitatively observed, LCF enabled monitoring of the oxidation process at different temperature steps. The analysis revealed an almost complete oxidation at 600\,°C after 15 minutes, with a component fraction of 97~$\pm$~5\,\% assigned to \ce{NiO/Mn2O3}.

\textit{Operando} measurements at the Ni K-edge on the 20-NiO/COK-12 catalyst provided detailed insight into the reduction and activation pathway, as well as the behavior under reaction conditions, with measurement time of 5 minutes per spectrum. The reduction/activation of NiO20-NiO/COK-12 began at approximately 450\,°C and progressed toward metallic Ni upon heating to 600\,°C, resulting in a complete edge shift toward the Ni metal reference. LCF indicated that at least 55\,\% of Ni was reduced, with the final fraction depending on both temperature and time. Importantly, temperature effects strongly influenced the analysis: elevated temperatures damped EXAFS oscillations via the Debye-Waller effect, biasing LCF fits toward oxide-like contributions. This explains the lower apparent metallic Ni fraction observed at 600\,°C ($\sim$66\,\%) compared to the higher fraction obtained after cooling ($\sim$91\,\%) in the \textit{in situ} experiment without GC analysis. These results emphasize the importance of acquiring reference spectra at elevated temperatures to improve the reliability of \textit{operando} LCF analysis, as done with the heating cell used in this work or with the cell developed by Praetz and Grötzsch \textit{et al.}\cite{Praetz-Groetzsch2025}
 
Under \ce{H2}/\ce{CO2} (4:1) reaction conditions, methane formation was confirmed by GC analysis at 350\,°C and 400\,°C, while the XAS spectra exhibited no significant structural changes beyond thermal damping and minor particle movement effects. 

The capabilities and limitations of the spectrometer setup in combination with the reactor cell using a quartz glass capillary were demonstrated. Employing a round quartz glass capillary -- which is advantageous for ensuring good gas flow through and around the particles, or for applying high pressure -- can, however, introduce artifacts in the spectra. These artifacts arise from the von Hámos geometry of the spectrometer used in this work, being most pronounced at lower energies and diminishing toward higher energies. Future developments in X-ray sources and optics, such as the implementation of pre-focusing optics to mitigate capillary-induced artifacts or micro-focused X-ray sources with higher efficiency at higher energies, could further increase the effectively usable spectral bandwidth and reduce measurement times while maintaining the same SNR. The addition of pressure-regulating equipment, coupled with a volumetric flow meter at the exit, coupled with MFCs intended for lower flows would lead to a significant increase of the precision in the gas composition analytics, allowing for much more complex analyses of the catalytic activity.  

In summary, this work demonstrates the feasibility of conducting \textit{in situ} and \textit{operando} XAS measurements with a laboratory-based von Hámos spectrometer, providing valuable insights into catalyst redox processes under realistic reaction conditions. Despite limitations from temperature effects -- also present in synchrotron experiments -- and capillary-induced artifacts, the results establish laboratory XAS as a powerful complementary tool to synchrotron experiments, enabling time-resolved studies with greater accessibility and flexibility. Future advances in reference measurements, reactor design, and spectrometer optics will further enhance data quality and reliability, opening new opportunities for routine \textit{operando} and \textit{in situ} investigations.

\section*{Author contributions}
\textbf{Sebastian Praetz}: Conceptualization, Investigation / Formal Analysis / Validation, Project administration, Resources (reference material), Software, Visualization, Writing - original draft, Writing - review and editing. \textbf{Emiliano Dal Molin}: Conceptualization, Investigation / Formal Analysis / Validation, Writing - original draft, Writing - review and editing. \textbf{Delf Kober}: Software, Resources, Investigation / Formal Analysis / Validation, Writing - review editing. \textbf{Marko Tesic}: Methodology, Software
(XAFS measurements). \textbf{Christopher Schlesiger}: Supervision, Validation, Software, Writing  - review and editing. \textbf{Peter Kraus}: Resources, Software, Writing - review and editing. \textbf{Julian T. Müller}: Resources, Writing - review and editing. \textbf{Aswin Jyothilakshmi Ravi}: Resources, Writing - review and editing. \textbf{Daniel Grötzsch}: Resources, Visualization, Writing – review and editing. \textbf{Maged F. Bekheet}: Supervision, Writing – review and editing. \textbf{Albert Gili}: Supervision, Writing – review and editing. \textbf{Aleksander Gurlo}: Conceptualization, Supervision, Funding acquisition, Writing – review and editing. \textbf{Birgit Kanngießer}: Supervision, Funding acquisition, Writing – review and editing

\section*{Conflicts of interest}
There are no conflicts to declare.

\section*{Data availability}
The data supporting this article have been included as part of the SI.
Furthermore, the raw and normalized XAS data for this article are available on Zenodo at \url{https://doi.org/10.5281/zenodo.17063731}.
The supplementary information includes: a detailed view of the von Hámos setup with the IR tube furnace reactor cell; XRD measurements of the synthesized 5\,\% Ni/MnO catalyst; raw and normalized data from the \textit{operando} and \textit{in situ} measurements of 20-NiO/COK-12; GC analyses; and initial comparative results obtained using the tungsten X-ray source optimized for 30 kV.

\section*{Acknowledgments}
This work was partially funded by the Deutsche Forschungsgemeinschaft (DFG German Research Foundation) under Germany's Excellence Strategy -- EXC2008-390540038 -- UniSysCat.
Part of this work was funded by the German Federal Ministry of Education and Research in the framework of the project Catlab  (03EW0015A).
Peter Kraus acknowledges funding from the DFG (Project No. 490703766).

\balance

\bibliography{references} 

\bibliographystyle{vancouver} 

\end{document}